\documentclass[aps, 10pt, prl, twocolumn, superscriptaddress, noshowpacs, preprintnumbers, floatfix]{revtex4-2}

\usepackage[colorlinks=true,breaklinks=true]{hyperref}
\hypersetup{allcolors=[rgb]{0.0 0.0 1.0},linkcolor=[rgb]{0.75 0.05 0.05}}
\usepackage[dvipsnames]{xcolor}
\usepackage{mathtools}
\usepackage{amsmath}
\usepackage{amsfonts}
\usepackage{amssymb}
\usepackage{bbold}
\usepackage{upgreek}
\usepackage{multirow}
\usepackage{bm}
\long\def\exclude#1{}
\usepackage{orcidlink}
\usepackage{slashed}

\newcommand{\beq}{\begin{equation}}
\newcommand{\eeq}{\end{equation}}
\newcommand{\bea}{\begin{eqnarray}}
\newcommand{\eea}{\end{eqnarray} }

\begin{document}
%TC:ignore
\title{Minimal Proton-Mass Dark Matter}

\author{Majed Khalaf \orcidlink{0000-0001-5537-9992}}
 \affiliation{Racah Institute of Physics, Hebrew University of Jerusalem, Jerusalem 91904, Israel}
\author{Eric Kuflik \orcidlink{0000-0003-0455-0467}}
 \affiliation{Racah Institute of Physics, Hebrew University of Jerusalem, Jerusalem 91904, Israel}%Lines break automatically or can be forced with \\
 \author{Alessandro Lenoci \orcidlink{0000-0002-2209-9262}}
 \affiliation{Racah Institute of Physics, Hebrew University of Jerusalem, Jerusalem 91904, Israel}
 \author{Hitoshi Murayama \orcidlink{0000-0001-5769-9471}}
 \affiliation{Leinweber Institute for Theoretical Physics, University of California, Berkeley, CA 94720, USA}
  \affiliation{Kavli Institute for the Physics and Mathematics of the Universe (WPI), University of Tokyo, Kashiwa 277-8583, Japan}
\affiliation{Theoretical Physics Group, Ernest Orlando Lawrence Berkeley National Laboratory, Berkeley, CA 94720, USA}
 \author{Edoardo~Vitagliano~\orcidlink{0000-0001-7847-1281}}
\affiliation{Dipartimento di Fisica e Astronomia, Universit\`{a} degli Studi di Padova, Via Marzolo 8, 35131 Padova, Italy}
\affiliation{Istituto Nazionale di Fisica Nucleare (INFN), Sezione di Padova, Via Marzolo 8, 35131 Padova, Italy}

\begin{abstract}
We present a minimal dark matter scenario: a single complex scalar carrying baryon and lepton number, with no new exact stabilizing symmetry. Its leading interaction is a dimension-7 semileptonic portal that, below confinement, generates a low-energy Yukawa coupling with the proton and electron. Requiring absolute stability of both the proton and dark matter forces the dark matter mass into a narrow window around the proton mass, which may be anthropically selected. Despite its minimal field content, the model can be probed by many observables: proton burning in stars, hydrogen decay, brown dwarfs and neutron star heating, and nucleon decay-like signatures in direct detection. UV-dominated freeze-in produces the observed relic abundance.  This framework provides a unique testable example of dark matter arising from a minimal extension of the Standard Model.
\end{abstract}
%TC:endignore

%\date{\today}

\maketitle

\begin{figure*}[t]
\includegraphics[width=0.8\textwidth]{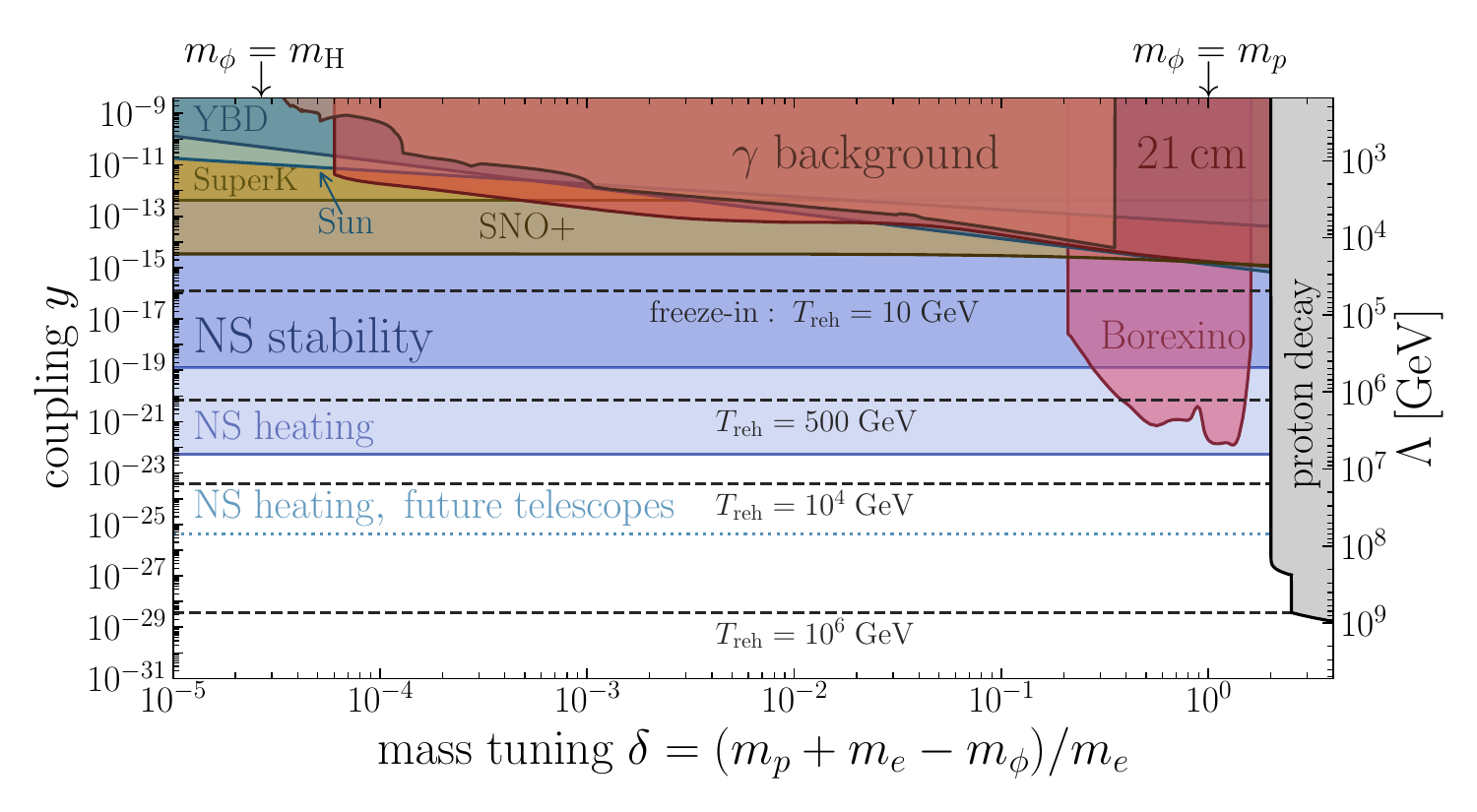}
\caption{ Constraints and projections in the $y,\delta$ plane. Red regions show hydrogen-decay bounds (21 cm, diffuse $\gamma$ background and Borexino constraints); blue regions show stellar (Sun and class-Y brown dwarfs, YBD) and neutron star constraints (stability and heating); yellow shows the indicative Super-Kamiokande~\cite{Super-Kamiokande:2002weg} sensitivity and brown shows the SNO+ invisible-neutron-decay constraint~\cite{SNO:2022trz}.  Dashed curves give the coupling required for UV freeze-in at the indicated reheating temperatures. The gray region is excluded by proton decay: the invisible-decay constrain~\cite{SNO:2022trz} and the $p\to e^+X$ bound~\cite{Super-Kamiokande:2002weg}. The dotted blue curve shows the projected heating sensitivity for a NS with a surface temperature of 1000 K. }
\label{fig:summary}
\end{figure*}

\textbf{\textit{Introduction}}---Dark matter is one of the few clear signals that the Standard Model is incomplete, yet its origin remains unknown~\cite{Planck:2018vyg}. The nature of dark matter (DM) is often addressed through hidden sectors, large extensions of the Standard Model (SM), or ad hoc stabilizing symmetries (see  reviews~\cite{Alexander:2016aln,Cooley:2022ufh,Cirelli:2024ssz}). In this work, we present a minimal model of DM, in which the SM is amended by a single complex scalar field $\phi$, with no new symmetries introduced; see  review~\cite{Cirelli:2024ssz} and references therein for other models of minimal DM.  
Our scalar DM candidate carries baryon and lepton number $B=L=-1$. We focus on the regime in which the interaction relevant for production and phenomenology is a dimension-7 portal inducing, after confinement, a Yukawa interaction $y \phi p e$. The observed abundance may be obtained from this interaction through ultraviolet freeze-in~\cite{Hall:2009bx,Elahi:2014fsa}. As displayed in Fig.~\ref{fig:summary}, the model can be tested through distinctive signatures such as hydrogen decay, stellar evolution, neutron stars, and nucleon destruction in the laboratory. We adopt natural units $\hbar=c=1$.

Our scalar DM $\phi$, taken to be an SM gauge singlet, interacts with SM fields via the dimension-7 portal
\beq
\frac{1}{\Lambda^3}\phi u u d e, \label{dim7operator}
\eeq
where $u$, $d$ and $e$ denote the right-handed up quark, down quark, and charged lepton. Here we suppressed Lorentz and gauge-index contractions for simplicity. For concreteness we focus on operators involving first generation fermions; other semi-leptonic portals of the equivalent form lead to similar low-energy phenomenology, though the precise thresholds depend on flavor.
 
After confinement, the operator induces effective interactions among hadrons, leptons and $\phi$. In particular, an effective Yukawa interaction of the form
\beq
y \phi \overline{p^c} P_R e \label{yukawaoperator}
\eeq
is generated, with $P_R$ the right chiral projector. Explicitly, we are considering the UV operator
\begin{equation}
\frac{1}{\Lambda^3}\phi\,\epsilon_{abc}
\, (u_R^{aT} C \,d_R^{b})(u_R^{cT} C \,e_R),  \qquad C\equiv i\gamma^2\gamma^0\ ,
\end{equation}
so the chiral structure of the Yukawa interaction is 
$
    y\, \phi\,p_R^T C e_R =y\, \phi\,\overline{p^c} P_R e 
$. 
From lattice computations of proton decay matrix elements~\cite{Aoki:2017puj}, the dimensionless Yukawa coupling can be estimated as $
y \simeq 0.0144~{\rm GeV}^3 /\Lambda^3 $, where $1/\Lambda^3$ is understood at the renormalization scale of 2 GeV. Most phenomenological signatures arise via this Yukawa interaction.

Given its quantum numbers, the DM field $\phi$ is stable if it is the lightest baryon or if kinematics forbid its decay into an antiproton and a positron. Absolute stability of both $\phi$ and the proton then requires 
\beq
m_p - m_e \le m_\phi \le m_p + m_e,  \label{stabledmp}
\eeq
which forbids both decays $\phi \to \bar{p}+e^+$ and $p \to \phi^*+e^+$~\footnote{The proton decay rate in our model is $$\tau^{-1} = \frac{y^2}{64\pi m_p^3}(m_p^2+m_e^2-m_\phi^2)\lambda^{1/2}(m_p^2,m_\phi,m_e),$$ where $\lambda(a,b,c)\equiv[a-(b+c)^2][a-(b-c)^2]$. Proton decay is constrained by visible decays $p\to e^+X$ in  Super-Kamiokande~\cite{Super-Kamiokande:2002weg} as $\tau> 7.9\times 10^{32}$ yr only if the positron has energy above the Cherenkov threshold $E_{\rm Ch}\simeq 0.775$ MeV. Otherwise, the bound by SNO+~\cite{SNO:2022trz} on invisible proton decay applies $\tau>0.96\times 10^{30}$ yr.}.
We discuss this tuning later.

Lower-dimensional portals involving $|\phi|^2$, such as $|\phi|^2|H|^2$ and dimension-6 operators, may also be present. Their phenomenology is well studied and is not the focus of this work. Instead, we focus on the unique physics of the dimension-7 coupling, coming from the low-energy Yukawa interaction in Eq.~\eqref{yukawaoperator}. Additional portal operators can modify the thermal history of $\phi$, but do not affect the phenomenology of the dimension-7 interaction considered here when the DM relic abundance is fixed.

The operator in Eq.~\eqref{dim7operator} might arise from simple renormalizable UV completions involving leptoquarks and diquarks. As one example, consider introducing the scalar fields with SM quantum numbers $S$~(${\bf 3},1,-1/3$)  and  $S'$~(${\bf \bar{3}},1,+1/3$). The relevant UV Lagrangian reads
\beq
\mathcal{L}_{\rm UV} = y_S S u d + y_{S'} S' u e+ g \phi^* S S'+ {\rm h.c.}
\eeq
Integrating out the heavy scalar fields at tree level generates the effective operator~\eqref{dim7operator}, with 
\beq
\frac{1}{\Lambda^3 } = \frac{ y_S y_{S'}g}{m_{S}^2 m_{S'}^2}\,. \label{LambdaUV}
\eeq
Baryon and lepton numbers are conserved and do not induce dangerous processes such as proton decay.

\textbf{\textit{Relic Abundance}}---Depending on the scale of $\Lambda$, the portal~\eqref{dim7operator} can lead to freeze-in~\cite{Hall:2009bx,Elahi:2014fsa}, symmetric freeze-out~\cite{Lee:1977ua}, or to an asymmetric/chemical-equilibrium scenario in which the $\phi$ abundance tracks the baryon abundance until late-time freeze-out. (For other DM models with freeze-out off a particle with a chemical potential see Refs.~\cite{Bandyopadhyay:2011qm,Farina:2016llk,Dror:2016rxc,Cline:2017tka,Berlin:2017ife,Kramer:2020sbb,Frumkin:2022ror}.)  

Freeze-out through assisted-decay~\cite{futurework} and hadronic channels requires at least $y > 10^{-9}$, excluded by stellar and terrestrial constraints. However, this scenario is rich and we explore it in the Supplemental Material (SupM). In the following, we focus on freeze-in.

We assume a negligible initial abundance of $\phi,\phi^*$ and standard radiation domination after reheating. The production is dominated by an ultraviolet (UV) freeze-in mediated by the operator~\eqref{dim7operator} \textit{prior} to the QCD phase transition. The case of production \textit{posterior} to QCD phase transition is studied in the SupM.

 The relevant processes are of the kind $f_1f_2 \to f_3f_4 \phi$ and $f_1f_2f_3 \to f_4\phi$, where $f_i=\{u,d,e\}$ denotes a SM fermion (two $u$ quarks are involved in all channels), along with the conjugate processes involving $\phi^*$. All fermions can be treated as massless and in thermal equilibrium with the plasma. 
 The total rate and the resulting relic abundance are calculated in the SupM. We obtain 
 \bea 
\frac{m_\phi Y_{\phi+\phi^{*}} }{ 0.44 ~{\rm eV} } 
&\simeq&  \left(\frac{y}{1.2 \times 10^{-21} }\right)^2 \left(\frac{T_{\rm reh}}{\rm 1~\rm TeV }\right)^5 \label{YFIprePT}
\eea
for reheating temperatures $T_{\rm reh} \gg m_p$. 
 
Since production is UV-dominated, our estimate assuming the contact-operator description remains valid up to $T_{\rm reh}$. 
In the low-energy effective theory this corresponds to requiring $T_{\rm reh} \lesssim \Lambda$, and in the sample UV completion $T_{\rm reh} \lesssim \min(m_S,m_{S'})$. From Eq.~\eqref{YFIprePT}, saturating the observed abundance implies
\beq
\frac{T_{\rm reh}}{\Lambda} \simeq \left(\frac{T_{\rm reh}}{1.3 \times 10^{23}\,{\rm GeV}}\right)^{1/6} \ll 1
\eeq
for any viable reheating temperature considered here. In terms of the UV completion in Eq.~\eqref{LambdaUV}, the condition $T_{\rm reh} \lesssim \min(m_S,m_{S'})$ is automatically satisfied provided the product $y_S y_{S'} g$ is not parametrically small compared to the mediator mass.

We considered the relic abundance being set solely by the interaction~\eqref{dim7operator}. Additional lower-dimension operators, such as $|\phi|^2 |H|^2$ may determine the relic abundance but do not affect the dimension-7 phenomenology considered below.

\textbf{\textit{Constraints from proton burning in stars}}---The stability of DM and the proton results in a bonanza of astrophysical and laboratory bounds, as well as detection perspectives, that we explore thoroughly.

In stellar interiors the coupling of Eq.~\eqref{yukawaoperator} allows
\beq\label{eq:H_burning}
  p +e^- \to  \gamma + \phi^*  ,
  % \qquad   p +\pi/\gamma  \to \phi^* + e^+,
\eeq 
throughout the stability window of Eq.~\eqref{stabledmp}. This process depletes the core of main-sequence stars of protons necessary for the nuclear reactions that sustain these stars against gravity. It is convenient to define the available energy release
$\Delta \equiv m_p + m_e - m_\phi$ with $\delta \equiv {\Delta}/{m_e}$, so that $0 \le \delta \le 2$ within Eq.~\eqref{stabledmp}; the channel is suppressed as $\delta \to 0$ since the final state becomes too heavy.

The lifetime of a Sun-like main-sequence star---i.e. the time it takes to burn the hydrogen in its core---is approximately $\tau_\odot\sim 7$ Gyr~\cite{Adelberger:2010qa,IAUInter-DivisionA-GWorkingGrouponNominalUnitsforStellarPlanetaryAstronomy:2015fjh,Vinyoles:2016djt}. This timescale is only approximate, since as the hydrogen content in the stellar core is depleted, the star expands into a red giant 
 and resorts to other nuclear reactions to sustain itself. 
The process in Eq.~\eqref{eq:H_burning} causes protons and electrons to be converted into $\phi^*$ and a photon, making them unavailable for the $pp$-chain, and altering stellar evolution. The proton-depletion rate is
\beq
\Gamma_{p\to\phi^*}=  n_e\,\langle\sigma v\rangle_{pe\to\phi^{*}\gamma}\simeq
n_e \frac{y^2\alpha_{\rm EM}\delta}{8 m_p\,m_e},
\label{eq:gamma_p_to_phi}
\eeq
with  $n_e\simeq 6\times 10^{25}\ {\rm cm}^{-3}$ the characteristic electron density in the Sun's core. Requiring that proton depletion does not compete with ordinary $pp$-burning over the main-sequence lifetime, $\Gamma_{p\to\phi^*}\,\tau_\odot \lesssim 1$, implies a bound
\beq 
y \lesssim \sqrt{ \frac{8m_p m_e}{\alpha_{\rm EM} \delta} \frac{1}{n_e \tau_\odot} }\simeq  \frac{6\times 10^{-14}}{\sqrt \delta} .
\eeq 
For larger couplings the proton-depletion process becomes too efficient, thereby shortening the Sun's lifetime, or, equivalently, reducing its present luminosity. 
A more robust bound can be obtained by solving the stellar evolution equations for the hydrogen fraction, including both the nuclear reaction rate and the process in Eq.~\eqref{eq:H_burning}. In a similar fashion, an even stronger constraint can be obtained from the neutrino luminosity in the Sun \cite{Gondolo:2008dd,Vinyoles:2015aba}.

\textbf{\textit{Constraints from brown dwarf heating}}---
Brown dwarfs (BDs) are objects not massive enough to sustain $pp$-burning, but emit light and heat via deuterium fusion. Remarkably, class-Y BDs have surface temperatures less than $500$ K \cite{Legget2017}.
The process \eqref{eq:H_burning} heats the core of brown dwarfs: for every reaction, $E_\gamma \simeq \delta m_e$ of energy is deposited in the BD. Hence, the heating rate is $H_{pe\to \phi^* \gamma}\simeq n_e \langle \sigma v\rangle_{pe \to \phi^* \gamma} \delta m_e N_p$ with $N_p\simeq 0.825 M_{\rm BD}/m_p$ the number of protons in the BD. We can estimate $n_e \simeq n_p\simeq N_p/V_{\rm BD}$, with $V_{\rm BD}$ the BD volume. Imposing $H< L_{\rm BD}$, with $L_{\rm BD}\simeq 4\pi \sigma_{\rm SB}R_{\rm BD}^2 T_{\rm BD}^4$ the Stefan-Boltzmann luminosity of a BD, we obtain a constraint 
\begin{align}
    y \lesssim \frac{10^{-15}}{\delta}\left( \frac{R_{\rm BD}}{0.1R_\odot} \right)^{5/2}\left( \frac{2\times 10^{-3}M_\odot}{M_{\rm BD}} \right)\left( \frac{T_{\rm BD}}{276\ {\rm K}} \right)^2\ .
\end{align}
We normalized the values to the ones for W0855, the coldest known class-Y BD,  having a luminosity of $6\times 10^{-8}L_\odot$ (see e.g. \cite{WISE1,coldestBD,Tinney2014,Luhman2014,Rowland2024,Legget2026}). Notice that a weaker bound can be obtained in the same fashion considering heating in the Sun, resulting in $y\lesssim 10^{-13}/\delta$.

\textbf{\textit{Constraints from neutron stars}}---Neutron stars (NS) provide strong constraints on any interaction that converts neutrons into bosons that can accumulate at low momentum. Such a bosonic population can soften the NS equation of state, since it does not benefit from Fermi degeneracy pressure, reducing the maximum supported mass. This is in tension with the observation of $2\, M_\odot$ neutron stars (see, e.g., Ref.~\cite{Baym:2018ljz,McKeen:2018xwc,Liu:2025qco,Divaris:2025ftf,Brugeat:2024rxe,Bastero-Gil:2024kjo,Husain:2022brl,Brandes:2023hma}).

Among the possibilities, the portal in Eq.~\eqref{yukawaoperator} induces the proton conversion processes
\begin{align}\label{eq:NS_process_22}
&p + e^- \to \gamma + \phi^*,\qquad \qquad  p + e^- \to \pi^0 + \phi^*,
\end{align}
Other channels involving a bystander nucleon contribute at the same order and are considered in the SupM.
At $\beta$-equilibrium, they all correspond to a subsequent neutron conversion, since a neutron will replace the missing proton. 
In a degenerate medium the kinematic threshold is controlled by the chemical potentials. The condition for the processes of Eq.~\eqref{eq:NS_process_22} to proceed is then $\mu_n=\mu_p + \mu_{e} \ge m_{\rm fin}$ with $m_{\rm fin}=m_\phi$ or $m_\phi+m_{\pi^0}$, which is satisfied in the inner core of sufficiently dense neutron stars for $m_\phi \sim m_p$. Conservatively, we neglected possible channels with charged pions, due to the large uncertainty on their mass~\cite{Fore:2023gwv}, which similarly affects axion emissivities in compact objects~\cite{Fiorillo:2025gnd}.

Once produced, the $\phi^*$ particles are mostly gravitationally trapped by the neutron star, since the escape velocity is $v_{\rm esc}=\sqrt{2GM_{\rm NS}/R_{\rm NS}}\simeq 0.7$, much larger than the average $\phi$ velocity for e.g. $M_{\rm NS}\simeq 2\, M_\odot$ and $R_{\rm NS}\simeq 11$ km. If production were efficient enough to approach chemical equilibrium, one would instead have $\mu_{\phi} = \mu_p + \mu_{e}$ and whenever $\mu_{\phi} > m_{\phi}$ the $\phi^*$ population would eventually form a Bose-Einstein condensate. In the absence of significant self-interactions, such a condensate is nearly pressure-less and tends to soften the equation of state. Possible self-interactions and medium-induced mixing may modify this conclusion, making the precise equilibrium regime model dependent.
Note that in our scenario thermalization does not occur since the $\phi$-nucleon scattering cross section is suppressed $\sigma_{\phi N}\propto y^4$.

We impose our constraint by requiring that nucleon conversion never removes more than 10\% of neutrons over the neutron star lifetime, $\tau_{\rm NS}\simeq 5\ {\rm Gyr}$. Let $\gamma(n_B)$ denote the total conversion rate per unit volume at baryon density $n_B$. This simple argument reads ${\gamma}/{n_n}\times\tau_{\rm NS} \lesssim 0.1$,
where $n_n$ is the neutron number density. We compute $\gamma(n_B)$ using the APR equation of state \cite{Akmal:1998cf} for $ n_B\simeq n_n = 0.6\ {\rm fm^{-3}}$,  corresponding to roughly 4 times the saturation density, which is about the core density expected for a NS of mass $2\, M_\odot$\cite{Brandes:2023hma}. An order-of-magnitude estimate of the bound can be obtained by
approximating $\gamma \simeq \sigma{(\Delta E)^6}/{(64\pi^4)}$. Here the denominator accounts for the phase space, the $s$-wave low-energy cross section is $\sigma\sim y^2\times 10^{-28}\ {\rm cm^2} $ for both processes in Eq.~\eqref{eq:NS_process_22}, and the energy difference between initial and final states is $\Delta E\simeq \mu_p+\mu_e-m_{\rm fin}\sim 300$ MeV, depending on the exact value of $n_B$ and $m_{\rm fin}$. This results in the constraint
\beq 
y \lesssim 2\times 10^{-19}\left( \frac{5\, {\rm Gyr}}{\tau_{\rm NS}}\right)^{1/2} \ .
\eeq
A careful numerical evaluation of all the processes (see SupM) with an integration over the radial density profile of the NS reveals that requiring chemical equilibrium to never be approached over the NS lifetime implies $y \lesssim 1.2\times 10^{-19}$, matching our order-of-magnitude estimate. 

An even stronger upper bound on the coupling derives from the observation of cooling NSs, as the processes~\eqref{eq:NS_process_22} can heat the star at late times~\cite{McKeen:2020oyr,McKeen:2021jbh}. Cold NSs older than $\sim 10^5\,\rm yr$ cool predominantly through photon emission from their surface with approximately blackbody luminosity $L_\gamma \simeq 4\pi R_{\rm NS}^2 \sigma_{\rm SB}T_{\rm NS}^4$, where $\sigma_{\rm SB}=\pi^2/60$ is the Stefan-Boltzmann constant, and $T_{\rm NS}$ the surface temperature. We can require that the energy injection from processes~\eqref{eq:NS_process_22} is compatible with the observed temperature of a cold NS. We estimate the heating rate per unit time as $H \simeq \sigma{(\Delta E)^7}V_{\rm NS}/{(64\pi^4 )}$ with $V_{\rm NS}\simeq M_{\rm NS}/(n_nm_n)$ the NS volume. Requiring  $H\lesssim L_\gamma$, we find
\begin{align}
    y \lesssim 10^{-23} \left( \frac{T_{\rm NS}}{3\times 10^4\ {\rm K}} \right) ^2 \left(\frac{ 1.5M_\odot}{M_{\rm NS}} \right) ^{1/2}\left( \frac{R_{\rm NS}}{ 11\ {\rm km}} \right),   
\end{align}
where we have chosen parameters compatible with the coldest known neutron star PSR J2144--3933 \cite{Guillot:2019ugf}. A more careful estimate of the heating rate, considering all the contributing processes and integrating over a radial density profile appropriate to PSR J2144--3933 (see SupM), results in the bound $y\lesssim 5.2\times 10^{-23}$, validating our simple estimate. Notice that the heating argument applies straightforwardly only if the coupling is small enough not to perturb the NS structure over the NS lifetime. For PSR J2144--3933, $\tau_{\rm NS}\simeq 0.3$ Gyr~\cite{Guillot:2019ugf} and this occurs when $y\lesssim 6\times 10^{-19}$. However, as seen above, larger couplings are  already excluded by older and heavier NS. Interestingly, the smallest coupling that can be probed by looking at cold NSs scales as $y\propto T_{\rm NS}^2$. Future surveys~\cite{LUVOIR, RubinLSST,DES,RomanWFIRST} could observe NSs whose surface temperature might be as low as 1000 K~\cite{McKeen:2021jbh}, corresponding to a sensitivity reach in our minimal model of down to $y\sim 4.3\times 10^{-26}$.

\textbf{\textit{Constraints from hydrogen decay}}---The possibility of hydrogen decay mediated by the Yukawa interaction in Eq.~\eqref{yukawaoperator} was considered by~Ref.~\cite{McKeen:2020zni}. The hydrogen lifetime for the decay ${\rm H} \to \phi +\gamma$ is estimated to be
\beq
\tau_{\rm H} \simeq 10^{32}~{\rm sec} \left(\frac{10^{-20}}{y}\right)^2 \delta^{-1},
\eeq 
and produces a monoenergetic photon $
E_\gamma^{\rm H}=
{(m_{\rm H}^2-m_\phi^2)}/{(2m_{\rm H})}$. Reference~\cite{McKeen:2020zni} concluded that  the search for the electron decay  $e \to \gamma +\textrm{missing energy}$  in Borexino~\cite{Borexino:2015qij} can be recast to constrain hydrogen decay, implying a hydrogen lifetime of order $10^{28}$ years. It should be noted that the Borexino detector is limited at low energies by intrinsic ${}^{14}{\rm C}$ backgrounds, losing sensitivity around $225$~keV. Therefore, if the photon energy $E_\gamma^{\rm H}$ falls below this value, the sensitivity to hydrogen decay is significantly reduced. A future improvement could come from JUNO, a $\sim$20-kton liquid-scintillator detector~\cite{JUNO:2021kxb,JUNO:2021vlw}. Naively, the increased target mass would enhance sensitivity relative to Borexino, although the scaling depends on whether the search is background-free. 
 
Neutral hydrogen is abundant in the universe, and decays that produce a photon 
are constrained by diffuse background observations. For a hydrogen lifetime $\tau_H$  the surface brightness from a direction $\hat n$ is
\beq
I_\gamma(\hat n)
  = \frac{1}{4\pi\tau_{\rm H}}\int_{\rm l.o.s.} ds \, n_{\rm H}(s,\hat n)
   \equiv \frac{1}{4\pi\tau_{\rm H}}\Sigma_{\rm H}(\hat n)\,,
\label{eq:Igamma_H}
\eeq
where $\Sigma_{\rm H}(\hat n)\equiv \int ds\,n_{\rm H}$ is the neutral-hydrogen column density along the line of sight. $\Sigma_{\rm H}$ depends strongly on the galactic latitude and longitude. We take the full-sky average value $\Sigma_H \simeq  10^{21}\ {\rm cm^{-2}}$~
\cite{Bekhti:2016fcs}. We use the diffuse background $I_\gamma^{\rm obs}$ brightness spectrum measurements collected in \cite{Porras-Bedmar:2024uql} as upper bounds. Conservatively, we require that $I_\gamma\lesssim I_\gamma^{\rm obs}$, so that
\beq
\tau_{\rm H}
\gtrsim \frac{\Sigma_{\rm H}}{4\pi I_\gamma^{\rm obs}}
\simeq 8\times 10^{19}\ {\rm s}
\left(\frac{\Sigma_{\rm H}}{10^{21} {\rm cm^{-2}}}\right)\!
\left(\frac{ {\rm cm^{-2}\,s^{-1}\, sr^{-1} }}{I_\gamma^{\rm obs}(E_\gamma^{\rm H})}\right)\!.
\label{eq:tauH_xray}
\eeq

Another constraint can be obtained by recasting the bounds on decaying relics derived from the sky-averaged 21 cm signal~\cite{Cima:2025zmc}. Observations of the hydrogen hyperfine transition are susceptible to additional injections of photons between recombination and the onset of star formation. Matching the energy-injection rates gives
\beq
\tau_{\rm H}
=
\tau_{\rm DM}(2E_\gamma^{\rm H})
\frac{\rho_{\rm H}}{2\rho_{\rm DM}}
\left(1-\frac{m_\phi^2}{m_{\rm H}^2}\right),
\label{eq:tauH_21cm}
\eeq
where $\tau_{\rm DM}(2E_\gamma^{\rm H})$ denotes the lifetime bound for a decaying particle with mass $m_{\rm DM}=2E_\gamma^{\rm H}$, so that the injected photon energy matches that of hydrogen decay. We take the cosmic ratio of hydrogen to DM mass densities to be
$\rho_{\rm H}/\rho_{\rm DM}\simeq 0.15$. We use the most conservative bounds on $\tau_{\rm DM}$ from Ref.~\cite{Cima:2025zmc} and show the resulting constrain in Fig~\ref{fig:summary}. Depending on the assumptions on the cosmological histories, the constraints on $\tau_{\rm H}$ could vary by about one order of magnitude.

\textbf{\textit{Direct detection}}---Dark matter in the galactic halo can scatter inside large detectors destroying a nucleon. The leading processes are $\phi + p/n \to e^++\pi$, mimicking baryon decay signatures, though with kinematics that differ from those expected in standard nucleon decay channels.
The total cross section in the non-relativistic limit is $
\sigma v  \simeq {3g_{\pi NN}^2 y^2}/({128\pi m_p^2})$ 
with $g_{\pi NN}\sim 13.5$. The corresponding scattering rate in a detector of mass $M$ reads
\bea
R = \frac{M}{m_p}\frac{\rho_{\rm DM}}{m_{\rm \phi}} \sigma v \simeq \frac{1}{\rm yr} \left( \frac{M}{22.5~\rm kton}\right) \left( \frac{y}{7 \times 10^{-13}}\right)^2 .
\eea
For reference, Super-Kamiokande has a fiducial mass of $22.5~\rm kton$~\cite{Super-Kamiokande:2002weg} and Hyper-Kamiokande will have $187~\rm kton$~\cite{Hyper-Kamiokande:2018ofw}. Fig.~\ref{fig:summary} shows an indicative event-per-year sensitivity. These events resemble nucleon decay but have a larger invariant mass near $m_\phi+m_p\sim 2m_p$, and a harder pion spectrum. Hence, existing searches would require a dedicated recast~\footnote{Similar signals from dark-matter-induced nucleon destruction have been discussed in Refs.~\cite{Davoudiasl:2011fj,Huang:2013xfa,Bell:2025uup}.}.

\textbf{\textit{Invisible neutron decay}}---The portal~\eqref{dim7operator} mediates the invisible neutron decay $n\to \phi^* +\overline \nu_e$ via a $W$ boson exchange. An estimate of the decay rate gives
\beq
\tau_n^{-1} \simeq \frac{y^2m_n}{32\pi}\left(1-\frac{m_\phi^2}{m_n^2}\right)^2 \left(\frac{2\sqrt2G_{\rm F} m_e m_u}{(4\pi)^2}\log\frac{M_W^2}{m_u^2}\right)^2,
\eeq
with $G_{\rm F}$ the Fermi constant, $m_u$ and $M_W$ the up-quark and $W$ boson mass, respectively.
Comparing it to the SNO+ constraint $\tau_n>\tau_{n\to {\rm inv}}=9\times 10^{29}\ {\rm yr}$~\cite{SNO:2022trz}, we get a bound of $y<10^{-15}$, varying by a factor of 2 between the allowed values of $m_\phi$.

\textbf{\textit{Anthropics}}---For the model to be viable, the dark matter mass must lie within the narrow range of Eq.~\eqref{stabledmp}, corresponding to a fine-tuning of order
\beq
\delta m_\phi^2 /m_\phi^2 =  2{\Delta m_\phi}/{m_\phi} \simeq  4 m_e/m_p\sim 2\times 10^{-3}
\,,
\eeq
where $\Delta m_\phi \simeq 2 m_e$.  
Although this is a large tuning, it can be subject to anthropic selection when the proton and electron masses are held fixed. If $m_\phi$ were even slightly heavier, the dark matter would become unstable and decay, eliminating the seeds of structure formation. Conversely, if $m_\phi$ were slightly lighter, the proton would decay and ordinary matter would not exist. Hence, observers could plausibly arise only in the narrow window where this tuning occurs.

Alternatively, one can view this as an anthropic tuning of the proton mass, while keeping $m_\phi$ fixed. Although the proton mass is dominated by the physics of QCD confinement, it still depends on the Higgs vacuum expectation value through the quark masses, which set thresholds in the QCD $\beta$-function and thereby affect the low-energy value of $\Lambda_{\rm QCD}$.  Therefore, for the same reasons as above: if the proton were marginally heavier or lighter, either matter or dark matter would disappear, respectively. Allowing both scalar masses to vary, one finds an anthropic degeneracy line near $m_p \simeq m_\phi$, along which observers can exist. It would be interesting to explore other scenarios where the tuning of the electroweak scale is tied to the existence of a DM relic.

\textbf{\textit{Conclusions and Outlook}}---We presented a DM scenario arising from a minimal extension of the Standard Model: a single complex scalar with feeble interactions and no new stabilizing symmetry, but requiring a strict mass window for the stability of DM and baryons. 
The strongest constraints arise from neutron stars, with additional sensitivity from hydrogen decay, stellar evolution, brown-dwarf heating and DM-induced nucleon destruction in large detectors. Future surveys able to observe NSs with surface temperatures as low as 1000 K could probe our minimal DM model down to $y\sim 10^{-26}$. It would be interesting to analyze the role of self-interactions and the formation of condensates in neutron stars, and to perform dedicated direct-detection searches for DM-induced nucleon destruction. 

In the minimal setup, the observed abundance is naturally obtained through UV-dominated freeze-in.  Although the case presented here does not yield a viable freeze-out realization, the assisted-decay and late baryon chemical-equilibrium mechanisms may suggest related freeze-out scenarios in other DM models~\cite{futurework}.

Finally, the unique anthropic feature of the model suggests exploring whether other DM candidates can be selected by similar mass conditions that ensure the survival of both DM and baryonic matter. 

 %TC:ignore
\textbf{\textit{Acknowledgments}}---We thank Csaba Cs\'{a}ki, Francesco D'Eramo, Yuval Grossman, Tom Hartman, Yonit Hochberg, Dmitry Ofengeim, Rotem Ovadia, Maxim Perelstein, Annachiara Picco, Maxim Pospelov, Georg Raffelt, Nir Shaviv, Nicholas Stone, and Ofri Telem for useful discussions. 

M.K., E.K. and A.L. are supported in part by grants No 2022713 and  2024091 from the US-Israel BSF/NSF and by grant No 2023711 from the US-Israel BSF. M.K., E.K. and A.L. are grateful to Cornell University for its hospitality and support during a sabbatical visit. M.K. and A.L. are grateful to the Azrieli Foundation for the award of
an Azrieli Fellowship. M.K. and A.L. are supported by an ERC STG grant (``Light-Dark,'' grant No. 101040019). H.M. is supported by the NSF grant PHY-2515115, by the U.S. Department of Energy (DE-AC02-05CH11231), by the JSPS Grant-in-Aid for Scientific Research JP23K03382, MEXT Grant-in-Aid for Transformative Research Areas (A) 26H00401, 26A204, 26H00403, Hamamatsu Photonics, K.K., Tokyo Dome Corporation, and by the World Premier International Research Center Initiative (WPI) MEXT, Japan. E.V. is supported by the Italian Ministero dell'Universit\`{a} e della Ricerca through Departments of Excellence grant 2023-2027 ``Quantum Frontier'', by the Italian Ministero dell'Universit\`{a} e della Ricerca through the FIS 2 project FIS-2023-01577 (DD n. 23314 10-12-2024, CUP C53C24001460001), and by Istituto Nazionale di Fisica Nucleare (INFN) through the
Theoretical Astroparticle Physics (TAsP) project. This project has received funding from the European Research Council (ERC) under the European Union’s Horizon Europe research and innovation programme (grant agreement No. 101040019). Views and opinions expressed are however those of the author(s) only and do not necessarily reflect those of the European Union. The European Union cannot be held responsible for them.

\bibliographystyle{biblio}
\bibliography{manref}

\providecommand{\href}[2]{#2}\begingroup\raggedright\begin{thebibliography}{10}

\bibitem{Planck:2018vyg}
{\scshape Planck} Collaboration, N.~Aghanim et~al., \emph{{Planck 2018 results.
  VI. Cosmological parameters}},
  \href{https://doi.org/10.1051/0004-6361/201833910}{\emph{Astron. Astrophys.}
  {\bfseries 641} (2020) A6}
  [\href{https://arxiv.org/abs/1807.06209}{{\ttfamily 1807.06209}}]. [Erratum:
  Astron.Astrophys. 652, C4 (2021)].

\bibitem{Alexander:2016aln}
J.~Alexander et~al., \emph{{Dark Sectors 2016 Workshop: Community Report}},  8,
  2016, \href{https://arxiv.org/abs/1608.08632}{{\ttfamily 1608.08632}}.

\bibitem{Cooley:2022ufh}
J.~Cooley et~al., \emph{{Report of the Topical Group on Particle Dark Matter
  for Snowmass 2021}},  \href{https://arxiv.org/abs/2209.07426}{{\ttfamily
  2209.07426}}.

\bibitem{Cirelli:2024ssz}
M.~Cirelli, A.~Strumia and J.~Zupan, \emph{{Dark Matter}},
  \href{https://arxiv.org/abs/2406.01705}{{\ttfamily 2406.01705}}.

\bibitem{Hall:2009bx}
L.~J. Hall, K.~Jedamzik, J.~March-Russell and S.~M. West, \emph{{Freeze-In
  Production of FIMP Dark Matter}},
  \href{https://doi.org/10.1007/JHEP03(2010)080}{\emph{JHEP} {\bfseries 03}
  (2010) 080} [\href{https://arxiv.org/abs/0911.1120}{{\ttfamily 0911.1120}}].

\bibitem{Elahi:2014fsa}
F.~Elahi, C.~Kolda and J.~Unwin, \emph{{UltraViolet Freeze-in}},
  \href{https://doi.org/10.1007/JHEP03(2015)048}{\emph{JHEP} {\bfseries 03}
  (2015) 048} [\href{https://arxiv.org/abs/1410.6157}{{\ttfamily 1410.6157}}].

\bibitem{Super-Kamiokande:2002weg}
{\scshape Super-Kamiokande} Collaboration, Y.~Fukuda et~al., \emph{{The
  Super-Kamiokande detector}},
  \href{https://doi.org/10.1016/S0168-9002(03)00425-X}{\emph{Nucl. Instrum.
  Meth. A} {\bfseries 501} (2003) 418}.

\bibitem{SNO:2022trz}
{\scshape SNO+} Collaboration, A.~Allega et~al., \emph{{Improved search for
  invisible modes of nucleon decay in water with the SNO+detector}},
  \href{https://doi.org/10.1103/PhysRevD.105.112012}{\emph{Phys. Rev. D}
  {\bfseries 105} (2022) 112012}
  [\href{https://arxiv.org/abs/2205.06400}{{\ttfamily 2205.06400}}].

\bibitem{Aoki:2017puj}
Y.~Aoki, T.~Izubuchi, E.~Shintani and A.~Soni, \emph{{Improved lattice
  computation of proton decay matrix elements}},
  \href{https://doi.org/10.1103/PhysRevD.96.014506}{\emph{Phys. Rev. D}
  {\bfseries 96} (2017) 014506}
  [\href{https://arxiv.org/abs/1705.01338}{{\ttfamily 1705.01338}}].

\bibitem{Note1}
The proton decay rate in our model is $$\tau ^{-1} = \protect \frac {y^2}{64\pi
  m_p^3}(m_p^2+m_e^2-m_\phi ^2)\lambda ^{1/2}(m_p^2,m_\phi ,m_e),$$ where
  $\lambda (a,b,c)\equiv [a-(b+c)^2][a-(b-c)^2]$. Proton decay is constrained
  by visible decays $p\to e^+X$ in Super-Kamiokande~\cite
  {Super-Kamiokande:2002weg} as $\tau > 7.9\times 10^{32}$ yr only if the
  positron has energy above the Cherenkov threshold $E_{\protect \rm Ch}\simeq
  0.775$ MeV. Otherwise, the bound by SNO+~\cite {SNO:2022trz} on invisible
  proton decay applies $\tau >0.96\times 10^{30}$ yr.

\bibitem{Lee:1977ua}
B.~W. Lee and S.~Weinberg, \emph{{Cosmological Lower Bound on Heavy Neutrino
  Masses}}, \href{https://doi.org/10.1103/PhysRevLett.39.165}{\emph{Phys. Rev.
  Lett.} {\bfseries 39} (1977) 165}.

\bibitem{Bandyopadhyay:2011qm}
P.~Bandyopadhyay, E.~J. Chun and J.-C. Park, \emph{{Right-handed sneutrino dark
  matter in $\mathbf{U(1)'}$ seesaw models and its signatures at the LHC}},
  \href{https://doi.org/10.1007/JHEP06(2011)129}{\emph{JHEP} {\bfseries 06}
  (2011) 129} [\href{https://arxiv.org/abs/1105.1652}{{\ttfamily 1105.1652}}].

\bibitem{Farina:2016llk}
M.~Farina, D.~Pappadopulo, J.~T. Ruderman and G.~Trevisan, \emph{{Phases of
  Cannibal Dark Matter}},
  \href{https://doi.org/10.1007/JHEP12(2016)039}{\emph{JHEP} {\bfseries 12}
  (2016) 039} [\href{https://arxiv.org/abs/1607.03108}{{\ttfamily
  1607.03108}}].

\bibitem{Dror:2016rxc}
J.~A. Dror, E.~Kuflik and W.~H. Ng, \emph{{Codecaying Dark Matter}},
  \href{https://doi.org/10.1103/PhysRevLett.117.211801}{\emph{Phys. Rev. Lett.}
  {\bfseries 117} (2016) 211801}
  [\href{https://arxiv.org/abs/1607.03110}{{\ttfamily 1607.03110}}].

\bibitem{Cline:2017tka}
J.~M. Cline, H.~Liu, T.~Slatyer and W.~Xue, \emph{{Enabling Forbidden Dark
  Matter}}, \href{https://doi.org/10.1103/PhysRevD.96.083521}{\emph{Phys. Rev.
  D} {\bfseries 96} (2017) 083521}
  [\href{https://arxiv.org/abs/1702.07716}{{\ttfamily 1702.07716}}].

\bibitem{Berlin:2017ife}
A.~Berlin, \emph{{WIMPs with GUTs: Dark Matter Coannihilation with a Lighter
  Species}}, \href{https://doi.org/10.1103/PhysRevLett.119.121801}{\emph{Phys.
  Rev. Lett.} {\bfseries 119} (2017) 121801}
  [\href{https://arxiv.org/abs/1704.08256}{{\ttfamily 1704.08256}}].

\bibitem{Kramer:2020sbb}
E.~D. Kramer, E.~Kuflik, N.~Levi, N.~J. Outmezguine and J.~T. Ruderman,
  \emph{{Heavy Thermal Dark Matter from a New Collision Mechanism}},
  \href{https://doi.org/10.1103/PhysRevLett.126.081802}{\emph{Phys. Rev. Lett.}
  {\bfseries 126} (2021) 081802}
  [\href{https://arxiv.org/abs/2003.04900}{{\ttfamily 2003.04900}}].

\bibitem{Frumkin:2022ror}
R.~Frumkin, E.~Kuflik, I.~Lavie and T.~Silverwater, \emph{{Roadmap to Thermal
  Dark Matter beyond the Weakly Interacting Dark Matter Unitarity Bound}},
  \href{https://doi.org/10.1103/PhysRevLett.130.171001}{\emph{Phys. Rev. Lett.}
  {\bfseries 130} (2023) 171001}
  [\href{https://arxiv.org/abs/2207.01635}{{\ttfamily 2207.01635}}].

\bibitem{futurework}
M.~Khalaf, E.~Kuflik, A.~Lenoci, H.~Murayama and E.~Vitagliano,
  \emph{{Freeze-out from Assisted Decays}}, {\emph{To appear} (2026) }.

\bibitem{Adelberger:2010qa}
E.~G. Adelberger et~al., \emph{{Solar fusion cross sections II: the pp chain
  and CNO cycles}}, \href{https://doi.org/10.1103/RevModPhys.83.195}{\emph{Rev.
  Mod. Phys.} {\bfseries 83} (2011) 195}
  [\href{https://arxiv.org/abs/1004.2318}{{\ttfamily 1004.2318}}].

\bibitem{IAUInter-DivisionA-GWorkingGrouponNominalUnitsforStellarPlanetaryAstronomy:2015fjh}
{\scshape IAU Inter-Division A-G Working Group on Nominal Units for Stellar
  {\&} Planetary Astronomy} Collaboration, E.~E. Mamajek et~al., \emph{{IAU
  2015 Resolution B3 on Recommended Nominal Conversion Constants for Selected
  Solar and Planetary Properties}},
  \href{https://arxiv.org/abs/1510.07674}{{\ttfamily 1510.07674}}.

\bibitem{Vinyoles:2016djt}
N.~Vinyoles, A.~M. Serenelli, F.~L. Villante, S.~Basu, J.~Bergstr{\"o}m, M.~C.
  Gonzalez-Garcia, M.~Maltoni, C.~Pe{\~n}a-Garay and N.~Song, \emph{{A new
  Generation of Standard Solar Models}},
  \href{https://doi.org/10.3847/1538-4357/835/2/202}{\emph{Astrophys. J.}
  {\bfseries 835} (2017) 202}
  [\href{https://arxiv.org/abs/1611.09867}{{\ttfamily 1611.09867}}].

\bibitem{Gondolo:2008dd}
P.~Gondolo and G.~G. Raffelt, \emph{{Solar neutrino limit on axions and
  keV-mass bosons}},
  \href{https://doi.org/10.1103/PhysRevD.79.107301}{\emph{Phys. Rev. D}
  {\bfseries 79} (2009) 107301}
  [\href{https://arxiv.org/abs/0807.2926}{{\ttfamily 0807.2926}}].

\bibitem{Vinyoles:2015aba}
N.~Vinyoles, A.~Serenelli, F.~L. Villante, S.~Basu, J.~Redondo and J.~Isern,
  \emph{{New axion and hidden photon constraints from a solar data global
  fit}}, \href{https://doi.org/10.1088/1475-7516/2015/10/015}{\emph{JCAP}
  {\bfseries 10} (2015) 015}
  [\href{https://arxiv.org/abs/1501.01639}{{\ttfamily 1501.01639}}].

\bibitem{Legget2017}
S.~K. {Leggett}, P.~{Tremblin}, T.~L. {Esplin}, K.~L. {Luhman} and C.~V.
  {Morley}, \emph{{The Y-type Brown Dwarfs: Estimates of Mass and Age from New
  Astrometry, Homogenized Photometry, and Near-infrared Spectroscopy}},
  \href{https://doi.org/10.3847/1538-4357/aa6fb5}{\emph{\apj} {\bfseries 842}
  (2017) 118} [\href{https://arxiv.org/abs/1704.03573}{{\ttfamily
  1704.03573}}].

\bibitem{WISE1}
E.~L. Wright et~al., \emph{{The Wide-field Infrared Survey Explorer (WISE):
  Mission Description and Initial On-orbit Performance}},
  \href{https://doi.org/10.1088/0004-6256/140/6/1868}{\emph{Astron. J.}
  {\bfseries 140} (2010) 1868}
  [\href{https://arxiv.org/abs/1008.0031}{{\ttfamily 1008.0031}}].

\bibitem{coldestBD}
J.~C. {Beam{\'\i}n}, V.~D. {Ivanov}, A.~{Bayo}, K.~{Mu{\v{z}}i{\'c}}, H.~M.~J.
  {Boffin}, F.~{Allard}, D.~{Homeier}, D.~{Minniti}, M.~{Gromadzki},
  R.~{Kurtev}, N.~{Lodieu}, E.~L. {Martin} and R.~A. {Mendez},
  \emph{{Temperature constraints on the coldest brown dwarf known: WISE
  0855-0714}},
  \href{https://doi.org/10.1051/0004-6361/201424505}{\emph{Astronomy and
  Astrophysics} {\bfseries 570} (2014) L8}
  [\href{https://arxiv.org/abs/1408.5424}{{\ttfamily 1408.5424}}].

\bibitem{Tinney2014}
C.~G. {Tinney}, J.~K. {Faherty}, J.~D. {Kirkpatrick}, M.~{Cushing}, C.~V.
  {Morley} and E.~L. {Wright}, \emph{{The Luminosities of the Coldest Brown
  Dwarfs}}, \href{https://doi.org/10.1088/0004-637X/796/1/39}{\emph{\apj}
  {\bfseries 796} (2014) 39} [\href{https://arxiv.org/abs/1410.0746}{{\ttfamily
  1410.0746}}].

\bibitem{Luhman2014}
K.~L. Luhman, \emph{Discovery of a~ 250 k brown dwarf at 2 pc from the sun},
  {\emph{The Astrophysical Journal Letters} {\bfseries 786} (2014) L18}.

\bibitem{Rowland2024}
M.~J. Rowland, C.~V. Morley, B.~E. Miles, G.~Suarez, J.~K. Faherty, A.~J.
  Skemer, S.~A. Beiler, M.~R. Line, G.~L. Bjoraker, J.~J. Fortney et~al.,
  \emph{Protosolar d-to-h abundance and one part per billion ph3 in the coldest
  brown dwarf}, {\emph{The Astrophysical Journal Letters} {\bfseries 977}
  (2024) L49}.

\bibitem{Legget2026}
S.~Leggett, \emph{The coldest known y dwarfs: Estimates of their effective
  temperatures}, {\emph{The Astrophysical Journal} {\bfseries 1002} (2026)
  113}.

\bibitem{Baym:2018ljz}
G.~Baym, D.~H. Beck, P.~Geltenbort and J.~Shelton, \emph{{Testing dark decays
  of baryons in neutron stars}},
  \href{https://doi.org/10.1103/PhysRevLett.121.061801}{\emph{Phys. Rev. Lett.}
  {\bfseries 121} (2018) 061801}
  [\href{https://arxiv.org/abs/1802.08282}{{\ttfamily 1802.08282}}].

\bibitem{McKeen:2018xwc}
D.~McKeen, A.~E. Nelson, S.~Reddy and D.~Zhou, \emph{{Neutron stars exclude
  light dark baryons}},
  \href{https://doi.org/10.1103/PhysRevLett.121.061802}{\emph{Phys. Rev. Lett.}
  {\bfseries 121} (2018) 061802}
  [\href{https://arxiv.org/abs/1802.08244}{{\ttfamily 1802.08244}}].

\bibitem{Liu:2025qco}
Y.~Liu, Z.~Liu, M.~Pospelov and S.~Reddy, \emph{{Constraints on Symmetric Dark
  Matter from Neutron Star Capture and Collapse}},
  \href{https://arxiv.org/abs/2508.04961}{{\ttfamily 2508.04961}}.

\bibitem{Divaris:2025ftf}
M.~Divaris and C.~C. Moustakidis, \emph{{Neutron Dark Decay in Neutron Stars:
  The Role of the Symmetry Energy}},
  \href{https://arxiv.org/abs/2508.21754}{{\ttfamily 2508.21754}}.

\bibitem{Brugeat:2024rxe}
T.~Brugeat and C.~Smith, \emph{{Dark-matter induced neutron-antineutron
  oscillations}}, \href{https://doi.org/10.1007/JHEP01(2025)132}{\emph{JHEP}
  {\bfseries 01} (2025) 132}
  [\href{https://arxiv.org/abs/2412.06434}{{\ttfamily 2412.06434}}].

\bibitem{Bastero-Gil:2024kjo}
M.~Bastero-Gil, T.~Huertas-Roldan and D.~Santos, \emph{{Neutron decay anomaly,
  neutron stars, and dark matter}},
  \href{https://doi.org/10.1103/PhysRevD.110.083003}{\emph{Phys. Rev. D}
  {\bfseries 110} (2024) 083003}
  [\href{https://arxiv.org/abs/2403.08666}{{\ttfamily 2403.08666}}].

\bibitem{Husain:2022brl}
W.~Husain and A.~W. Thomas, \emph{{Novel neutron decay mode inside neutron
  stars}}, \href{https://doi.org/10.1088/1361-6471/aca1d5}{\emph{J. Phys. G}
  {\bfseries 50} (2023) 015202}
  [\href{https://arxiv.org/abs/2206.11262}{{\ttfamily 2206.11262}}].

\bibitem{Brandes:2023hma}
L.~Brandes, W.~Weise and N.~Kaiser, \emph{{Evidence against a strong
  first-order phase transition in neutron star cores: Impact of new data}},
  \href{https://doi.org/10.1103/PhysRevD.108.094014}{\emph{Phys. Rev. D}
  {\bfseries 108} (2023) 094014}
  [\href{https://arxiv.org/abs/2306.06218}{{\ttfamily 2306.06218}}].

\bibitem{Fore:2023gwv}
B.~Fore, N.~Kaiser, S.~Reddy and N.~C. Warrington, \emph{{Mass of charged pions
  in neutron-star matter}},
  \href{https://doi.org/10.1103/PhysRevC.110.025803}{\emph{Phys. Rev. C}
  {\bfseries 110} (2024) 025803}
  [\href{https://arxiv.org/abs/2301.07226}{{\ttfamily 2301.07226}}].

\bibitem{Fiorillo:2025gnd}
D.~F.~G. Fiorillo, {\'A}.~Gil~Muyor, H.-T. Janka, G.~G. Raffelt and
  E.~Vitagliano, \emph{{Axion-photon conversion in transient compact stars:
  Systematics, constraints, and opportunities}},
  \href{https://doi.org/10.1088/1475-7516/2026/03/053}{\emph{JCAP} {\bfseries
  03} (2026) 053} [\href{https://arxiv.org/abs/2509.13322}{{\ttfamily
  2509.13322}}].

\bibitem{Akmal:1998cf}
A.~Akmal, V.~R. Pandharipande and D.~G. Ravenhall, \emph{{The Equation of state
  of nucleon matter and neutron star structure}},
  \href{https://doi.org/10.1103/PhysRevC.58.1804}{\emph{Phys. Rev. C}
  {\bfseries 58} (1998) 1804}
  [\href{https://arxiv.org/abs/nucl-th/9804027}{{\ttfamily nucl-th/9804027}}].

\bibitem{McKeen:2020oyr}
D.~McKeen, M.~Pospelov and N.~Raj, \emph{{Cosmological and astrophysical probes
  of dark baryons}},
  \href{https://doi.org/10.1103/PhysRevD.103.115002}{\emph{Phys. Rev. D}
  {\bfseries 103} (2021) 115002}
  [\href{https://arxiv.org/abs/2012.09865}{{\ttfamily 2012.09865}}].

\bibitem{McKeen:2021jbh}
D.~McKeen, M.~Pospelov and N.~Raj, \emph{{Neutron Star Internal Heating
  Constraints on Mirror Matter}},
  \href{https://doi.org/10.1103/PhysRevLett.127.061805}{\emph{Phys. Rev. Lett.}
  {\bfseries 127} (2021) 061805}
  [\href{https://arxiv.org/abs/2105.09951}{{\ttfamily 2105.09951}}].

\bibitem{Guillot:2019ugf}
S.~Guillot, G.~G. Pavlov, C.~Reyes, A.~Reisenegger, L.~Rodriguez, B.~Rangelov
  and O.~Kargaltsev, \emph{{Hubble Space Telescope Nondetection of PSR
  J2144{\textendash}3933: The Coldest Known Neutron Star}},
  \href{https://doi.org/10.3847/1538-4357/ab0f38}{\emph{Astrophys. J.}
  {\bfseries 874} (2019) 175}
  [\href{https://arxiv.org/abs/1901.07998}{{\ttfamily 1901.07998}}].

\bibitem{LUVOIR}
{The LUVOIR Team}, \emph{{The LUVOIR Mission Concept Study Final Report}},
  \href{https://doi.org/10.48550/arXiv.1912.06219}{\emph{arXiv e-prints} (2019)
  arXiv:1912.06219} [\href{https://arxiv.org/abs/1912.06219}{{\ttfamily
  1912.06219}}].

\bibitem{RubinLSST}
J.~R. Peterson, \emph{Dark energy studies with lsst image simulations, final
  report},  tech. rep., Purdue Univ., West Lafayette, IN (United States), 07,
  2016.
\newblock 10.2172/1272167.

\bibitem{DES}
H.~T. Diehl, \emph{The dark energy survey and operations: Year 6 – the
  finale},  tech. rep., SLAC National Accelerator Laboratory (SLAC), Menlo
  Park, CA (United States); Fermi National Accelerator Laboratory (FNAL),
  Batavia, IL (United States), 01, 2020.
\newblock 10.2172/1596042.

\bibitem{RomanWFIRST}
J.~{Green} et~al., \emph{{Wide-Field InfraRed Survey Telescope (WFIRST) Final
  Report}}, \href{https://doi.org/10.48550/arXiv.1208.4012}{\emph{arXiv
  e-prints} (2012) arXiv:1208.4012}
  [\href{https://arxiv.org/abs/1208.4012}{{\ttfamily 1208.4012}}].

\bibitem{McKeen:2020zni}
D.~McKeen and M.~Pospelov, \emph{{How Long Does the Hydrogen Atom Live?}},
  \href{https://doi.org/10.3390/universe9110473}{\emph{Universe} {\bfseries 9}
  (2023) 473} [\href{https://arxiv.org/abs/2003.02270}{{\ttfamily
  2003.02270}}].

\bibitem{Borexino:2015qij}
{\scshape Borexino} Collaboration, M.~Agostini et~al., \emph{{A test of
  electric charge conservation with Borexino}},
  \href{https://doi.org/10.1103/PhysRevLett.115.231802}{\emph{Phys. Rev. Lett.}
  {\bfseries 115} (2015) 231802}
  [\href{https://arxiv.org/abs/1509.01223}{{\ttfamily 1509.01223}}].

\bibitem{JUNO:2021kxb}
{\scshape JUNO} Collaboration, A.~Abusleme et~al., \emph{{Radioactivity control
  strategy for the JUNO detector}},
  \href{https://doi.org/10.1007/JHEP11(2021)102}{\emph{JHEP} {\bfseries 11}
  (2021) 102} [\href{https://arxiv.org/abs/2107.03669}{{\ttfamily
  2107.03669}}].

\bibitem{JUNO:2021vlw}
{\scshape JUNO} Collaboration, A.~Abusleme et~al., \emph{{JUNO physics and
  detector}}, \href{https://doi.org/10.1016/j.ppnp.2021.103927}{\emph{Prog.
  Part. Nucl. Phys.} {\bfseries 123} (2022) 103927}
  [\href{https://arxiv.org/abs/2104.02565}{{\ttfamily 2104.02565}}].

\bibitem{Bekhti:2016fcs}
N.~B. Bekhti et~al., \emph{{HI4PI: a full-sky H{\,}i survey based on EBHIS and
  GASS}}, \href{https://doi.org/10.1051/0004-6361/201629178}{\emph{Astron.
  Astrophys.} {\bfseries 594} (2016) }
  [\href{https://arxiv.org/abs/1610.06175}{{\ttfamily 1610.06175}}].

\bibitem{Porras-Bedmar:2024uql}
S.~Porras-Bedmar, M.~Meyer and D.~Horns, \emph{{Novel bounds on decaying
  axionlike particle dark matter from the cosmic background}},
  \href{https://doi.org/10.1103/PhysRevD.110.103501}{\emph{Phys. Rev. D}
  {\bfseries 110} (2024) 103501}
  [\href{https://arxiv.org/abs/2407.10618}{{\ttfamily 2407.10618}}].

\bibitem{Cima:2025zmc}
F.~Cima and F.~D'Eramo, \emph{{Probing non-minimal dark sectors via the 21 cm
  line at cosmic dawn}},
  \href{https://doi.org/10.1088/1475-7516/2026/02/020}{\emph{JCAP} {\bfseries
  02} (2026) 020} [\href{https://arxiv.org/abs/2507.10664}{{\ttfamily
  2507.10664}}].

\bibitem{Hyper-Kamiokande:2018ofw}
{\scshape Hyper-Kamiokande} Collaboration, K.~Abe et~al.,
  \emph{{Hyper-Kamiokande Design Report}},
  \href{https://arxiv.org/abs/1805.04163}{{\ttfamily 1805.04163}}.

\bibitem{Note2}
Similar signals from dark-matter-induced nucleon destruction have been
  discussed in Refs.~\cite {Davoudiasl:2011fj,Huang:2013xfa,Bell:2025uup}.

\bibitem{Davoudiasl:2011fj}
H.~Davoudiasl, D.~E. Morrissey, K.~Sigurdson and S.~Tulin, \emph{{Baryon
  Destruction by Asymmetric Dark Matter}},
  \href{https://doi.org/10.1103/PhysRevD.84.096008}{\emph{Phys. Rev. D}
  {\bfseries 84} (2011) 096008}
  [\href{https://arxiv.org/abs/1106.4320}{{\ttfamily 1106.4320}}].

\bibitem{Huang:2013xfa}
J.~Huang and Y.~Zhao, \emph{{Dark Matter Induced Nucleon Decay: Model and
  Signatures}}, \href{https://doi.org/10.1007/JHEP02(2014)077}{\emph{JHEP}
  {\bfseries 02} (2014) 077} [\href{https://arxiv.org/abs/1312.0011}{{\ttfamily
  1312.0011}}].

\bibitem{Bell:2025uup}
N.~F. Bell, P.~Cox, J.~L. Newstead and M.~B.~G. Verde, \emph{{Dark matter
  induced nucleon decay through the neutron portal}},
  \href{https://doi.org/10.1103/8qfr-v4wx}{\emph{Phys. Rev. D} {\bfseries 113}
  (2026) 055033} [\href{https://arxiv.org/abs/2511.18722}{{\ttfamily
  2511.18722}}].

\bibitem{Peccei:1968bct}
R.~D. Peccei, \emph{{Chiral lagrangian calculation of pion-nucleon scattering
  lengths}}, \href{https://doi.org/10.1103/PhysRev.176.1812}{\emph{Phys. Rev.}
  {\bfseries 176} (1968) 1812}.

\bibitem{Pich1995}
A.~Pich, \emph{Chiral perturbation theory},
  \href{https://doi.org/10.1088/0034-4885/58/6/001}{\emph{Reports on Progress
  in Physics} {\bfseries 58} (1995) 563–609}.

\bibitem{Epelbaum:2015vea}
E.~Epelbaum, J.~Gegelia, U.-G. Mei{\ss}ner and D.-L. Yao, \emph{{Baryon chiral
  perturbation theory extended beyond the low-energy region}},
  \href{https://doi.org/10.1140/epjc/s10052-015-3728-7}{\emph{Eur. Phys. J. C}
  {\bfseries 75} (2015) 499}
  [\href{https://arxiv.org/abs/1510.02388}{{\ttfamily 1510.02388}}].

\bibitem{Bhattacharya:2014ara}
T.~Bhattacharya et~al., \emph{{QCD Phase Transition with Chiral Quarks and
  Physical Quark Masses}},
  \href{https://doi.org/10.1103/PhysRevLett.113.082001}{\emph{Phys. Rev. Lett.}
  {\bfseries 113} (2014) 082001}
  [\href{https://arxiv.org/abs/1402.5175}{{\ttfamily 1402.5175}}].

\bibitem{Laine:2015kra}
M.~Laine and M.~Meyer, \emph{{Standard Model thermodynamics across the
  electroweak crossover}},
  \href{https://doi.org/10.1088/1475-7516/2015/07/035}{\emph{JCAP} {\bfseries
  07} (2015) 035} [\href{https://arxiv.org/abs/1503.04935}{{\ttfamily
  1503.04935}}].

\bibitem{NSCool}
D.~{Page}, \emph{{NSCool: Neutron star cooling code}},  Astrophysics Source
  Code Library, record ascl:1609.009, Sept., 2016.

\bibitem{Brandes:2023bob}
L.~Brandes and W.~Weise, \emph{{Constraints on Phase Transitions in Neutron
  Star Matter}}, \href{https://doi.org/10.3390/sym16010111}{\emph{Symmetry}
  {\bfseries 16} (2024) 111}
  [\href{https://arxiv.org/abs/2312.11937}{{\ttfamily 2312.11937}}].

\bibitem{Berryman:2023rmh}
J.~M. Berryman, S.~Gardner and M.~Zakeri, \emph{{How macroscopic limits on
  neutron-star baryon loss yield microscopic limits on non-standard-model
  baryon decay}},
  \href{https://doi.org/10.1103/PhysRevD.109.023021}{\emph{Phys. Rev. D}
  {\bfseries 109} (2024) 023021}
  [\href{https://arxiv.org/abs/2305.13377}{{\ttfamily 2305.13377}}].

\bibitem{Friman:1979ecl}
B.~L. Friman and O.~V. Maxwell, \emph{{Neutron Star Neutrino Emissivities}},
  \href{https://doi.org/10.1086/157313}{\emph{Astrophys. J.} {\bfseries 232}
  (1979) 541}.

\bibitem{Bottaro:2024ugp}
S.~Bottaro, A.~Caputo and D.~F.~G. Fiorillo, \emph{{Neutrino emission in cold
  neutron stars: Bremsstrahlung and modified urca rates reexamined}},
  \href{https://doi.org/10.1088/1475-7516/2024/11/015}{\emph{JCAP} {\bfseries
  11} (2024) 015} [\href{https://arxiv.org/abs/2406.18640}{{\ttfamily
  2406.18640}}].

\bibitem{Fiorillo:2025zzx}
D.~F.~G. Fiorillo, A.~Lella, C.~A.~J. O'Hare and E.~Vitagliano, \emph{{Leading
  Bounds on Micrometer to Picometer Fifth Forces from Neutron Star Cooling}},
  \href{https://doi.org/10.1103/tlqz-713s}{\emph{Phys. Rev. Lett.} {\bfseries
  135} (2025) 211003} [\href{https://arxiv.org/abs/2506.19906}{{\ttfamily
  2506.19906}}].

\bibitem{NANOGrav:2019jur}
{\scshape NANOGrav} Collaboration, H.~T. Cromartie et~al., \emph{{Relativistic
  Shapiro delay measurements of an extremely massive millisecond pulsar}},
  \href{https://doi.org/10.1038/s41550-019-0880-2}{\emph{Nature Astron.}
  {\bfseries 4} (2019) 72} [\href{https://arxiv.org/abs/1904.06759}{{\ttfamily
  1904.06759}}].

\bibitem{Fonseca:2021wxt}
E.~Fonseca et~al., \emph{{Refined Mass and Geometric Measurements of the
  High-mass PSR J0740+6620}},
  \href{https://doi.org/10.3847/2041-8213/ac03b8}{\emph{Astrophys. J. Lett.}
  {\bfseries 915} (2021) L12}
  [\href{https://arxiv.org/abs/2104.00880}{{\ttfamily 2104.00880}}].

\bibitem{Riley:2021pdl}
T.~E. Riley et~al., \emph{{A NICER View of the Massive Pulsar PSR J0740+6620
  Informed by Radio Timing and XMM-Newton Spectroscopy}},
  \href{https://doi.org/10.3847/2041-8213/ac0a81}{\emph{Astrophys. J. Lett.}
  {\bfseries 918} (2021) L27}
  [\href{https://arxiv.org/abs/2105.06980}{{\ttfamily 2105.06980}}].

\bibitem{Gonzalez:2010ta}
D.~Gonzalez and A.~Reisenegger, \emph{{Internal Heating of Old Neutron Stars:
  Contrasting Different Mechanisms}},
  \href{https://doi.org/10.1051/0004-6361/201015084}{\emph{Astron. Astrophys.}
  {\bfseries 522} (2010) A16}
  [\href{https://arxiv.org/abs/1005.5699}{{\ttfamily 1005.5699}}].

\bibitem{Iwamoto:1984ir}
N.~Iwamoto, \emph{{Axion Emission from Neutron Stars}},
  \href{https://doi.org/10.1103/PhysRevLett.53.1198}{\emph{Phys. Rev. Lett.}
  {\bfseries 53} (1984) 1198}.

\bibitem{Fiorillo:2026dqu}
D.~F.~G. Fiorillo, A.~Lella, G.~G. Raffelt, N.~Selimovic and E.~Vitagliano,
  \emph{{Production of Leptophilic Bosons in Ultradegenerate Relativistic
  Matter}},  \href{https://arxiv.org/abs/2605.24081}{{\ttfamily 2605.24081}}.

\bibitem{Fiorillo:2026wso}
D.~F.~G. Fiorillo, A.~Lella, G.~G. Raffelt, N.~Selimovic and E.~Vitagliano,
  \emph{{Neutron Star Bounds on Muonic Fifth Forces from Picometer to Kilometer
  Scales}},  \href{https://arxiv.org/abs/2605.24094}{{\ttfamily 2605.24094}}.

\end{thebibliography}\endgroup
\onecolumngrid

\onecolumngrid
% \appendix

\setcounter{equation}{0}
\setcounter{figure}{0}
\setcounter{table}{0}
\setcounter{page}{1}
\makeatletter
\renewcommand{\theequation}{S\arabic{equation}}
\renewcommand{\thefigure}{S\arabic{figure}}
\renewcommand{\thepage}{S\arabic{page}}

\begin{center}
\textbf{\large Supplemental Material for the Letter: {\em Minimal Proton-Mass Dark Matter}}
\end{center}

\bigskip
In this Supplemental Material, we provide results complementing the content of the main text. In Section~A, we collect amplitudes, cross-sections and rates which recur in the dark matter (DM) relic density calculation and other phenomenology. We study in detail the freeze-in production of DM in Section~B, where we distinguish between production after the QCD phase transition (QCDPT) and prior to it, and we conclude that the latter is the dominant production channel. In Section~C, we provide more details on the freeze-out DM production, which represents an interesting benchmark, despite the required coupling being excluded by several observations. Section~D focuses on neutron stars (NS), which are the source of the most stringent constraints on the $\phi pe$ portal. Finally, Section~E shows how to compute general  collisional integrals using Monte Carlo techniques.
\bigskip
\section{A.~Binary scatterings}
Here we collect results for the various binary scatterings considered in the Supplemental Material and in the Letter.
\subsection{I.~Useful definitions}
We first introduce some notation for a generic scattering $12\leftrightarrow34$. The rate per unit volume is given by
\beq\label{rate1234}
\gamma_{12\leftrightarrow34}  =  \int  d\Pi_1d\Pi_2d\Pi_3d\Pi_4\, (2\pi)^4\delta^{(4)}(P_1+P_2-P_3-P_4)|{\cal M}_{12\to 34}|^2f_1^{\rm eq}f_2^{\rm eq}\,.
\eeq
Here, $P_i=(E_i,\vec{p}_i)$ are four-momenta, and $d\Pi_i= g_id^3p_i/[(2\pi)^3(2E_i)]$ is the Lorentz-invariant phase-space element, including the internal degrees of freedom $g_i$ of particle $i$. $|{\cal M}_{12\to34}|^2$ denotes the squared matrix element of the process, averaged over all internal degrees of freedom, and is expressible using the Mandelstam variables $s=(P_1+P_2)^2$ and $t=(P_1-P_3)^2 $. Throughout this section, we use Maxwell-Boltzmann (MB) statistics, neglect quantum-statistical factors $(1\pm f)$, and assume CP conservation so that forward and backward matrix elements are equal. 

The rate is related to thermally averaged cross-sections via
\beq \label{conversion_ann_prod}
\gamma_{12\leftrightarrow34}=\left< \sigma v \right>_{12\to34}n_1^{\rm eq}n_2^{\rm eq} = \left< \sigma v \right>_{34\to12}n_3^{\rm eq}n_4^{\rm eq}\,,
\eeq  
where, in the MB approximation the number densities are
\begin{equation}
    n_i^{\rm eq}= g_i\int \frac{d^3p}{(2\pi)^3}\, f_i^{\rm eq}(p) = \begin{dcases}
         \frac{g_i}{\pi^2}T^3 & T\gg m_i\,{\rm or}\,m_i=0\\
        \frac{g_i}{2\pi^2}T^3 \left( \frac{m_i}{T} \right)^2 K_{2} \left[\frac{m_i}{T}\right] &T\lesssim m_i
    \end{dcases}\,.
\end{equation}

The rate can be computed numerically once the squared amplitude of the process is known. Importantly, the rate can be reduced to a simpler integral over two variables. We start by using the 4-dimensional Dirac delta to integrate over ${\bf p}_4$ and $E_3$, obtaining the form
\begin{align}
   \gamma_{12\leftrightarrow34}=  \int d\Pi_1 d\Pi_2 [2 \sqrt{\lambda(s,m_1,m_2)}  \sigma_{12\to 34}(s)] f_1^{\rm eq} f_2^{\rm eq}\,.
\end{align}
Here $\lambda(a,b,c)\equiv[a-(b+c)^2][a-(b-c)^2]$ is the K\"{a}ll\'{e}n function and the cross section is defined as
\beq \sigma_{12\to 34}(s) = \frac{g_3 g_4}{16\pi \lambda (s,m_1,m_2)} \int_{t_-}^{t_+} dt\,|{\cal M}_{12\to 34}|^2 \,.
\eeq
Simplifying the two-body initial phase space, we can write the rate as 
\begin{align}
    \gamma_{12\to 34}(T) \equiv \frac{g_1g_2g_3 g_4 T}{1024 \pi^5}   \int_{s_-}^\infty \frac{ds}{s^{3/2}} \sqrt{\lambda(s,m_1,m_2)}\sqrt{\lambda(s,m_3,m_4)}  K_{1}[\sqrt{s}/T]\int_{-1}^{1} d\cos \theta |{\cal M}_{12\to 34}|^2.
\end{align}
with $s_-=\max[(m_1+m_2)^2,(m_3+m_4)^2]$ and $K_1$ the Bessel-$K$ function of first kind. It is useful to rewrite the rate in the compact form
\begin{align}\label{rate_1234_final}
    \gamma_{12\leftrightarrow 34}(T) \equiv \frac{g_1g_2g_3 g_4 T^4}{1024 \pi^5}   {\cal D}  (T,m_1,m_2,m_3,m_4)\,,
\end{align}
where we define the dimensionless function
\begin{align}
    {\cal D}(T,m_1,m_2,m_3,m_4)\equiv \int_{\max[(x_1+x_2)^2,(x_3+x_4)^2]}^\infty \frac{d\xi}{\xi^{3/2}} \sqrt{\lambda(\xi,x_1,x_2)}\sqrt{\lambda(\xi,x_3,x_4)}  K_{1}[ \sqrt{\xi}]\int_{-1}^{1} d\cos \theta |{\cal M}_{12\to 34}|^2 \,
\end{align}
with $x_i=m_i/T$. Note how $\gamma_{12\leftrightarrow 34}$ is explicitly symmetric under the exchange of initial and final states. 
When the matrix element is expressed as a function of $s,t$, the $t$ variable has to be replaced by
\begin{align}
    t(s,\cos\theta)= m_1^2 +m_3^2 - \frac{1}{2s}(s+m_1^2-m_2^2)(s+m_3^2-m_4^2) + \frac{\cos\theta}{2s}\sqrt{\lambda(s,m_1,m_2)}\sqrt{\lambda(s,m_3,m_4)}.
\end{align}
Unless specified otherwise, we use Eq.~\eqref{rate_1234_final} to numerically compute rates when three out of four particles are at equilibrium with the early universe plasma.

\subsection{II.~Matrix elements, rates and cross sections}
Below, we list key quantities for all the binary scatterings considered in this work. The low energy Lagrangian below the QCDPT, where binary scatterings are relevant, reads
\begin{align}\label{IRlagrangian}
 {\cal L}_{T<T_{\rm PT}} = {\cal L}_\phi + {\cal L}_{\rm EM}+{\cal L}_\pi
 \end{align} 
where
 \begin{subequations}
  \begin{align}\label{eq:L_phi}
 {\cal L}_\phi=&\,\partial_\mu \phi^* \partial^\mu \phi - m_\phi^2 \phi^* \phi + y \phi \overline{p^c} P_R e + {\rm h.c.} \\ \label{eq:L_QED}
{\cal L}_{\rm EM}=&\, -\frac{1}{4}F_{\mu\nu}F^{\mu\nu} 
+\overline e[i\gamma^\mu (\partial_\mu - ie A_\mu)- m_e]e 
+ \overline p[i\gamma^\mu (\partial_\mu + ie A_\mu)- m_p]p\\ \label{eq:L_pi}
{\cal L}_{\pi}=&\, \partial_\mu \pi^+ \partial^\mu \pi^- - m_{\pi^\pm}^2 \pi^+\pi^- + \frac{1}{2} \partial_\mu \pi^0 \partial^\mu {\pi^0} - \frac{1}{2}m_{\pi^0}^2 {\pi^0}^2 -\frac{g_A}{2F_\pi}\partial_\mu \pi^0 (\overline p \gamma^\mu \gamma^5p-\overline n \gamma^\mu \gamma^5n) \nonumber\\
&- \frac{g_A}{\sqrt{2}F_\pi}\partial_\mu \pi^+  \overline p \gamma^\mu \gamma^5n + \frac{i}{4 F^2_\pi}(\pi^+\partial_\mu \pi^-  -\pi^-\partial_\mu \pi^+) \left(\overline p \gamma^\mu p - \overline n \gamma^\mu n\right) + \frac{i\sqrt{2}}{4 F^2_\pi}(\pi^0\partial_\mu \pi^+  -\pi^+\partial_\mu \pi^0) \overline p \gamma^\mu n + {\rm h.c.}
\end{align}
\end{subequations}
This Lagrangian consists of the effective $pe\phi$ interaction and kinetic terms for our beyond the Standard Model complex scalar (first line), the EM Lagrangian and the chiral Lagrangian (see e.g. \cite{Peccei:1968bct,Pich1995,Epelbaum:2015vea}). The matter content is the following: fermions include the electron $e$, the proton $p$ and the neutron $n$ Dirac fields; bosons include the photon $A_\mu$, the DM complex scalar field $\phi$ and the pions (neutral $\pi^0$ and charged $\pi^\pm$).
In the following $\delta=(m_p+m_e-m_\phi) /m_e$ is employed.

\begin{itemize}
\item {$p+\gamma\leftrightarrow \phi+e$}: The squared amplitude averaged over all degrees of freedom reads:
\begin{align}\nonumber\label{ampl:p_gamma_phi_e}
   |{\cal M}_{p+\gamma\to \phi+e}|^2=& \frac{e^2 y^2}{4}   \bigg[-\frac{2 \left(s(s+t)+m_{\phi }^2 \left(m_e^2+3 m_p^2-s-t\right)+m_e^2 \left(s+t-3 m_p^2-m_e^2\right)+m_p^2(m_p^2-3s-t)\right)}{(m_p^2-s) (m_p^2+m_{\phi }^2-s-t)}\\\nonumber
&\qquad\qquad+\frac{s(s+t)+m_p^2(m_p^2-4s-t-2m_e^2)+m_{\phi }^2 \left(3 m_p^2-s\right)}{(m_p^2-s){}^2}\\
   &\qquad\qquad+\frac{s (s+t)+2 m_e^2 \left(s+t-2 m_p^2-m_e^2\right)+m_p^2 (m_p^2-2 s-t)+m_{\phi }^2 \left(m_p^2-s\right)}{(m_p^2+m_{\phi }^2-s-t)^2}\bigg]\,.
    \end{align}
Analytical expressions for the rates can be obtained in two different low temperature limits at leading order:
        \begin{subequations}
       \begin{align}
     \gamma_{p\gamma\leftrightarrow \phi e}^{m_e \ll T\ll m_p}\simeq & \frac{y^2 \alpha_{\rm EM}}{(2\pi)^{7/2}}m_p^{1/2}T^{7/2}e^{-m_p/T} \log\frac{2T}{m_e}%- \gamma_{\rm E} +1
     \,,\\
     \gamma_{p\gamma\leftrightarrow \phi e}^{T\ll m_e}\simeq &\frac{y^2 \alpha_{\rm EM}}{16\pi^3}(m_pm_e)^{1/2} T^3e^{-\frac{m_p+m_e(2-\delta)}{T}}(2-\delta)\,.
     \end{align}
    \end{subequations}
    From this, thermally averaged cross sections can be obtained. Relevant ones are
      \begin{subequations}
        \begin{align}
     \langle \sigma v \rangle_{p\gamma \to \phi e}^{m_e\ll T\ll m_p}\simeq &\frac{y^2 \alpha_{\rm EM}}{{16}m_p T} \log\frac{2T}{m_e} , %-\gamma_{\rm E} +1
     \,\\
     \langle \sigma v \rangle_{\phi e \to p\gamma}^{m_e\ll T\ll m_p}\simeq &\frac{y^2 \alpha_{\rm EM}}{{8}m_p T} \log\frac{2T}{m_e}%- \gamma_{\rm E} +1
     \,,\\
     \langle \sigma v \rangle_{ p \gamma \to\phi e}^{T\ll m_e}  \simeq&\frac{y^2 \alpha_{\rm EM}}{{16}m_pm_e}(2-\delta) \sqrt{\frac{\pi m_e^3}{2 T^3}} e^{-(2-\delta) m_e /T},\\
     \langle \sigma v \rangle_{\phi e \to p \gamma}^{T\ll m_e}  \simeq&\frac{y^2 \alpha_{\rm EM}}{{4}m_p m_e}(2-\delta).
    \end{align}
    \end{subequations}
    
    \item {$p+e\leftrightarrow \phi+\gamma$}: The squared amplitude averaged over all degrees of freedom can be obtained from Eq.~\eqref{ampl:p_gamma_phi_e} via the following crossing symmetry:
    \begin{align}\label{ampl:p_e_gamma_phi}
    |{\cal M}_{p+e\to \phi+\gamma}(s',t')|^2 = -|{\cal M}_{p+\gamma\to \phi+e}(s\to t',t\to s')|^2 \,.
    \end{align}    
Analytical expressions for the rates can be obtained in two different low temperature limits at leading order:
    \begin{subequations}
       \begin{align}
     \gamma_{pe\leftrightarrow \phi\gamma}^{m_e\ll T\ll m_p}\simeq & \frac{y^2 \alpha_{\rm EM}}{(2\pi)^{7/2}}m_p^{1/2}T^{7/2}e^{-m_p/T} \log\frac{2T}{m_e}
    \,,\\
     \gamma_{pe\leftrightarrow \phi\gamma}^{T\ll m_e}\simeq & \frac{y^2 \alpha_{\rm EM}}{16\pi^3}(m_pm_e)^{1/2} T^3 e^{-\frac{m_p+m_e}{T}}\delta\,.
     \end{align}
    \end{subequations}
    From this, thermally averaged cross sections can be obtained. Relevant ones are
    \begin{subequations}
        \begin{align}
     \langle \sigma v \rangle_{pe\to \phi\gamma}^{m_e\ll T\ll m_p}\simeq &\frac{y^2 \alpha_{\rm EM}}{{16}m_p T} \log\frac{2T}{m_e}%- \gamma_{\rm E} +1
    \,,\\
     \langle \sigma v \rangle_{\phi\gamma\to pe}^{m_e\ll T\ll m_p}\simeq & 
\frac{y^2 \alpha_{\rm EM}}{{8}m_p T} \log\frac{2T}{m_e}
\,, \\
     \langle \sigma v \rangle_{pe\to \phi\gamma}^{T\ll m_e}  \simeq&\frac{y^2 \alpha_{\rm EM}}{{8}m_p m_e}\delta,\\
          \langle \sigma v \rangle_{ \phi\gamma \to pe}^{T\ll m_e}  \simeq&\frac{y^2 \alpha_{\rm EM}}{{8}m_pm_e}\delta \sqrt{\frac{\pi m_e^3}{2 T^3}} e^{-\delta m_e /T}.
    \end{align}
    \end{subequations}
\item $e+\gamma \leftrightarrow \phi + p $: Similarly, applying crossing symmetry to Eq.~\eqref{ampl:p_gamma_phi_e} yields
\begin{align}
  |{\cal M}_{e+\gamma \leftrightarrow \phi + p}(s',t')|^2 =  |{\cal M}_{p+\gamma \leftrightarrow \phi + e}(s\to u', t\to t')|^2,
\end{align}
where $u' = m_e^2+m_{\phi}^2+m_{p}^2-s'-t'$. In the low temperature limit $T\ll m_p$, the rate is given by
\beq
\gamma^{T\ll m_p}_{e\gamma\leftrightarrow \phi p} \simeq \frac{y^2\alpha_{\rm EM}  T^3 m_p}{64 \pi ^3} e^{-\frac{2 m_p}{T}}\,,
\eeq
with the corresponding thermally averaged cross sections
\begin{subequations}
    \begin{align}
      \langle \sigma v \rangle_{\phi p\to e \gamma}^{T\ll m_p}\simeq &\frac{y^2 \alpha_{\rm EM} }{16 m_p^2}
    \,,\\
     \langle \sigma v \rangle_{e\gamma\to \phi p}^{T\ll m_p}\simeq & 
     \frac{\pi  y^2 \alpha _{\text{EM}} m_p}{128 T m_e^2
   K_2\left(\frac{m_e}{T}\right)} e^{-\frac{2 m_p}{T}} 
     \,.   
    \end{align}
    \end{subequations}
\item {$N+e\leftrightarrow \phi+\pi$}: For these hadronic processes the matrix element averaged over all degrees of freedom is 
    \begin{align}\label{eq:M_pe_phipi}
        |{\cal M}_{N+e \leftrightarrow \phi + \pi}|^2 =  
       \frac{g_{\pi NN}^2y^2 f_N^2}{16 \left(m_p^2-t\right)^2} \bigg[& st^2-t^2-2st + 4 m_\pi^2 t- 4m_\phi^2 m_\pi^2 + m_p^2 (s+2t) -m_p^4 - m_e^2 t^2 - m_e^2 m_p^2 + 2 m_e^2 (2m_\pi^2 +t)\bigg].
    \end{align}
Here, $N=p,n$ such that  $(m_N,m_\pi,f_N)$ are either $(m_p,m_{\pi^0},1)$ or $(m_n,m_{\pi^\pm},\sqrt{2})$ depending on whether the external state is a proton or a neutron, respectively. Assuming $m_\phi \simeq m_p$, the scattering rate in the $m_e \ll T \ll m_\pi$ temperature range and at next-to-next-to-leading order in $m_\pi/m_p$ is given by
\beq 
\gamma^{m_e\ll T\ll m_\pi}_{\phi \pi \leftrightarrow N e}\simeq \frac{g_{\pi NN}^2 y^2 f_N^2 m_{\pi }^{9/2} T^3 }{1024 \pi ^4 m_p^{7/2}} e^{\frac{-m_p-m_{\pi }}{T}} \left( 1-\frac{7}{2} \frac{m_{\pi }}{m_p}+\frac{63}{8} \frac{m_{\pi }^2}{m_p^2}\right)\,.
\eeq 
where we took $m_N\simeq m_p$ for simplicity. From the rate we can obtain the thermally averaged cross sections:
\bea
\left<\sigma v \right>^{m_e\ll T\ll m_\pi}_{ \phi\pi \to N e} &\simeq&\frac{y^2 g_{\pi NN}^2f_N^2 m_\pi^3}{128 \pi {m_p^5}}\left( 1-\frac{7}{2} \frac{m_{\pi }}{m_p}+\frac{63}{8} \frac{m_{\pi }^2}{m_p^2}\right), \\
\left< \sigma v \right>^{m_e\ll T\ll m_\pi}_{ N e \to \phi\pi} &\simeq &\frac{y^2 g_{\pi NN}^2 f_N^2 m_{\pi }^{9/2}}{1024 \sqrt{2 \pi } T^{3/2} m_p^5} e^{-\frac{m_{\pi }}{T}} \left( 1-\frac{7}{2} \frac{m_{\pi }}{m_p}  +\frac{63}{8} \frac{m_{\pi }^2}{m_p^2}\right).
\eea

\item {$N+\pi\leftrightarrow \phi+e$}: We obtain the fully averaged matrix element from Eq.~\eqref{eq:M_pe_phipi} using crossing symmetry
\begin{align}
  |{\cal M}_{N+\pi \leftrightarrow \phi + e}(s',t')|^2 =  -  |{\cal M}_{N+e \leftrightarrow \phi + \pi}(s\to t', t\to s')|^2.
\end{align}
Under the assumption $m_\phi \simeq m_p \simeq m_n$, the scattering rate in the $m_e \ll T \ll m_\pi$ temperature range and at next-to-next-to-leading order in $m_\pi/m_p$ is given by
\beq 
\gamma^{m_e\ll T\ll m_\pi}_{\phi e \leftrightarrow N \pi} \simeq \frac{g_{\pi NN}^2 y^2 f_N^2 m_{\pi }^{9/2} T^3 }{1024 \pi ^4 m_p^{7/2}} e^{\frac{-m_p-m_{\pi }}{T}} \left( 1-\frac{9}{2} \frac{m_{\pi }}{m_p}+\frac{99}{8} \frac{m_{\pi }^2}{m_p^2}\right)\,.
\eeq 
The following thermally-averaged cross sections then read:
\bea
\left<\sigma v \right>^{m_e\ll T\ll m_\pi}_{ \phi e \to N \pi} &\simeq& \frac{y^2 g_{\pi NN}^2f_N^2 m_\pi^{9/2}}{512 \sqrt{2\pi} {m_p^8} T^{3/2}}\left( 1-\frac{9}{2} \frac{m_{\pi }}{m_p}+\frac{99}{8} \frac{m_{\pi }^2}{m_p^2}\right),\\
\left< \sigma v \right>^{T\ll m_\pi}_{ N \pi \to \phi e} &\simeq&\frac{y^2 g_{\pi NN}^2f_N^2 m_\pi^3}{256 \pi {m_p^5}}\left( 1-\frac{9}{2} \frac{m_{\pi }}{m_p}+\frac{99}{8} \frac{m_{\pi }^2}{m_p^2}\right).
\eea
\item {$\pi+e\leftrightarrow \phi+N$}: We obtain the fully averaged matrix element from Eq.~\eqref{eq:M_pe_phipi} using crossing symmetry
\begin{align}
  |{\cal M}_{\pi+e \leftrightarrow \phi + N}(s',t')|^2 =  -  |{\cal M}_{N+e \leftrightarrow \phi + \pi}(s\to u', t\to t')|^2,
\end{align}
\end{itemize}
where $u' = m_{\pi}^2+m_e^2+m_{\phi}^2+m_{N}^2-s'-t'$. In the limit $
T\ll m_p$, the rate is given by
\beq
\gamma^{T\ll m_p}_{e\pi\leftrightarrow \phi N} \simeq \frac{y^2 g_{\pi N N}^2 f_N^2 T^3  m_p}{512 \pi ^4} e^{-\frac{2 m_p}{T}}\,,
\eeq
with the corresponding thermally averaged cross sections
\begin{subequations}
    \begin{align}
      \langle \sigma v \rangle_{\phi N\to e \pi}^{T\ll m_p}\simeq &\frac{y^2 g_{\pi N N}^2 f_N^2 }{128 \pi m_p^2}
    \,,\\
     \langle \sigma v \rangle_{e\pi\to \phi N}^{T\ll m_p}\simeq & 
     \frac{y^2 g_{\pi N N}^2 f_N^2 m_p T}{256 m_e^2 m_\pi^2
   K_2\left(\frac{m_e}{T}\right) K_2\left(\frac{m_\pi}{T}\right)} e^{-\frac{2 m_p}{T}} 
     \,.   
    \end{align}
    \end{subequations}

\section{B.~Dark Matter Freeze-in}
In this section, we review the dark matter freeze-in production. We distinguish between two cases: the dark matter produced via freeze-in (i)  after QCDPT, via binary scatterings involving hadrons or (ii) before QCDPT, via scatterings of quarks. We will see how UV freeze-in prior to the QCDPT is the dominant dark matter production channel if reheating occurs prior to the QCDPT.

\subsection{I.~After the QCDPT}
We begin with the freeze-in production immediately after the QCD phase transition from the Yukawa interaction $y \phi \overline{p^c}P_R e$. The production channels for $\phi^*$ via binary scatterings are electromagnetic and pionic channels
\begin{subequations} \label{app_fichannels}
\bea 
  &p +e^- \to \phi^* +  \gamma, \qquad \qquad & p+ \gamma  \to \phi^* + e^+,  \\
  &p +e^- \to \phi^* + \pi^0,  \qquad \qquad &  p +\pi^0   \to \phi^* + e^+, \\
  &n+e^- \to \phi^* + \pi^-,  \qquad \qquad &  n +\pi^+   \to \phi^* + e^+,
\eea 
\end{subequations}
and analogously for $\phi$. Several $3\to 2$ and $2\to3$ processes involving two pions also contribute. The processes can be grouped in three categories, with channels in each category related by crossing symmetry: (i) channels with a charged and a neutral pion, (ii) two charged pions, (iii) two neutral pions: 
\begin{subequations}
\begin{align}
\pi^+ +n +\pi^0 &\rightarrow \phi^* + e^+ \qquad + \quad {\rm crossing\ symmetry,}\\
\pi^+ +p +\pi^- &\rightarrow \phi^* + e^+\qquad + \quad {\rm crossing\ symmetry,} \\
\pi^0 +p +\pi^0 &\rightarrow \phi^* + e^+\qquad + \quad {\rm crossing\ symmetry.}
\end{align}
\end{subequations}

The $\pi^\pm\pi^0$ and $\pi^\pm\pi^\mp$ channels involve 10 processes, with 4 of them having final state mass larger than the initial state mass. The $\pi^0\pi^0$
channels include 7 processes with 3 of them suppressed by the final state mass. Due to the larger initial or final state mass with respect to binary scatterings, and hadrons and pions being Boltzmann-suppressed immediately after the QCD phase transition, collectively all the processes involving two pions matter only at the highest temperatures at the same level as the binary scattering $n+e^-\to \phi^*+\pi^-$. Following this reasoning, we neglect all processes except binary scatterings.

The total rate for $\phi^*$ production is
\beq
\gamma_{\phi}^{T<T_{\rm PT}}(T) =  \sum_{i,j,k} \gamma_{ij\leftrightarrow k\phi}\,. 
\eeq
The sum runs over the channels in Eq.~\eqref{app_fichannels} using the definition of $\gamma_{12\leftrightarrow34}$ as in Eq.~\eqref{rate1234}. 
We assumed all the particles but $\phi^*$ are in equilibrium throughout freeze-in production (freeze-in ends before nucleon-antinucleon annihilation). 

The hadronic processes with electrons in the initial state dominate the rate. The processes involving neutrons and protons are equivalent at the percent level up to a relative factor of 2.

\begin{figure}
    \centering
    \includegraphics[width=0.7\linewidth]{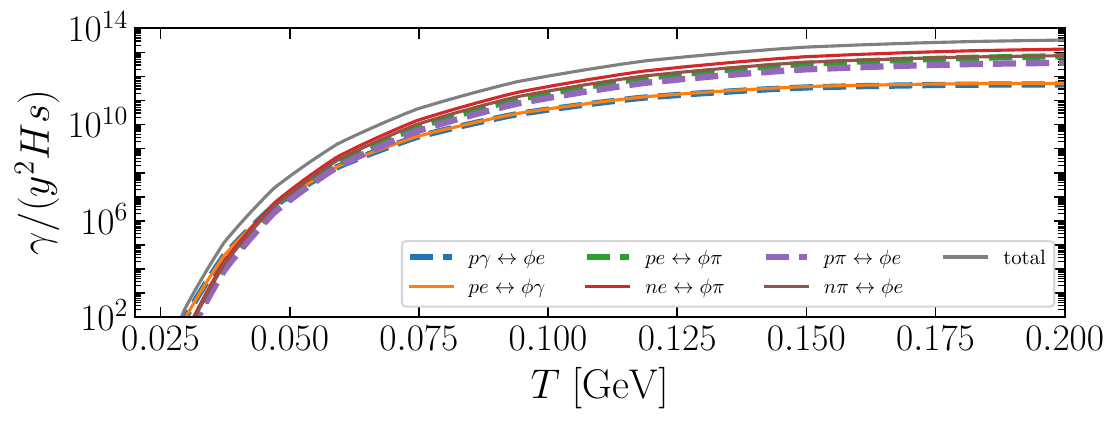}
    \caption{The interaction rates for the considered production channels, normalized by the coupling squared, Hubble parameter and entropy density as a function of temperature just below the QCD phase transition. The gray line shows the total rate. }
    \label{fig:rate}
\end{figure}

In Fig.~\ref{fig:rate}, we show the single-process rates as a function of temperature, the total rate and its approximations.
The total DM yield $Y_{\phi+\phi^*}=Y_\phi+Y_{\phi^*}$ can be obtained via direct integration of the Boltzmann equation starting from the QCD phase transition temperature, $T_{\rm start}= 155$~MeV~\cite{Bhattacharya:2014ara} and multiplying by a factor of 2 to take into account the production of both $\phi$ and $\phi^*$:
\bea
Y_{\phi+\phi^*}(T_0)= -2\int^{T_{\rm start}}_{T_0} \frac{dT}{T}\frac{\gamma_\phi^{T<T_{\rm PT}}(T)}{s(T) H(T)}\left(1+ \frac{1}{3}\frac{d\log g_{*s}}{d\log T} \right)\,.
\eea 
In the above equation $s(T)=2\pi^2g_{*s}(T)T^3/45 $ is the entropy density of the universe, $H(T)=T^2\sqrt{\pi^2 g_{*\rho}(T)/90}/M_{\rm Pl} $ is the Hubble parameter during radiation domination and $(g_{*\rho},g_{*s})$ are the number of relativistic degrees of freedom in energy density and entropy \cite{Laine:2015kra}. $M_{\rm Pl}$ is the reduced Planck mass. The observed relic abundance $m_\phi Y_{\phi+\phi^*}(T_0) = 0.44$ eV is obtained for $y = 6.9 \times 10^{-12}$.

One can verify that for this value of the coupling $y$ the production rate at $T=T_{\rm start}$ is much less than the Hubble rate
\bea 
\Gamma_\phi^{T<T_{\rm PT}}(T_{\rm start}) = \frac{\gamma_\phi^{T<T_{\rm PT}}(T_{\rm start})}{n_\phi^{\rm eq}(T_{\rm start})} \ll H(T_{\rm start}) \,,
\eea
confirming the validity of freeze-in.

\subsection{II.~Before the QCDPT}
Before the QCD phase transition the DM abundance is set via UV-dominated freeze-in through the dimension-7 operator in Eq.~\eqref{dim7operator}. At high temperatures, the relevant Standard Model degrees of freedom are electrons and quarks. Restoring the gauge and Lorentz structure of the term, we have the following operator to be added to the quark Lagrangian,
\begin{align}
    {\cal L}_{T\gg T_{\rm PT}} \supset \frac{\phi}{\Lambda^3}\epsilon_{abc}({d^{b}_R}^T i\gamma^0\gamma^2 u_R^a)
    ({e_R}^T i\gamma^0\gamma^2 u_R^c) + {\rm h.c.}
\end{align}
mediating a 5-point contact interaction between four fermions and $\phi$. In this calculation, we can safely assume that all the involved particles are massless, since we will consider production at $T\gg {\rm GeV}$. The DM is produced through the  following $2\to 3$ processes  
\beq
    uu \to  \bar d e^+ \phi^*,
    \quad
    d e^- \to  \bar u \bar u \phi^*,
    \quad
    u  e^-  \to \bar u \bar d \phi^*,
    \quad
    u d  \to  \bar u e^+ \phi^*\,, 
\eeq
and the following $3\to 2$ processes
\beq
    uud \to   e^+ \phi^*,\quad
     uue^- \to  \bar d \phi^*,\quad
  ude^-  \to \bar u \phi^* \,.
\eeq
with the addition of all the conjugate processes involving $\phi$ in the final state. We define the expression

\begin{align}
    |{\cal M}_5(i,j,k,\ell)|^2 = &\frac{1}{18 \Lambda^6}[2(P_i \cdot P_k) (P_j \cdot P_\ell)+2 (P_i \cdot P_\ell) (P_j \cdot P_k) -(P_i \cdot P_j) (P_k \cdot P_\ell)  ]  ,
\end{align}
which is symmetric under swaps $i\leftrightarrow j$ and $k\leftrightarrow \ell$. We use crossing symmetries to relate the squared matrix elements of the different channels.  For the processes of the type $12\to345$ we obtain the following values for the matrix elements averaged over all the internal degrees of freedom
\begin{align}
   |{\cal M}_{uu\to de \phi}|^2 &=|{\cal M}_5(1,2,3,4)|^2,  \\
    |{\cal M}_{de\to uu \phi}|^2 &=|{\cal M}_5(3,4,1,2)|^2 , \\
        |{\cal M}_{ue\to ud \phi}|^2 &=|{\cal M}_5(1,4,3,2)|^2 , \\
         |{\cal M}_{ud\to ue \phi}|^2 &=|{\cal M}_5(1,3,2,4)|^2, 
\end{align}
while for 
processes of the type $123\to45$ we obtain
\begin{align}
   |{\cal M}_{uud\to  e \phi}|^2 &=|{\cal M}_5(1,2,3,4)|^2 , \\
    |{\cal M}_{uue\to d\phi}|^2 &=|{\cal M}_5(1,2,4,3)|^2 ,  \\
|{\cal M}_{ude\to u\phi}|^2 &=|{\cal M}_5(1,3,4,2)|^2  .
\end{align}

To compute the total rate we decompose it in rates of the form $\gamma_{n\to m}$ where $n+m=5$ and $n,m$ indicate the number of particles in the initial and final state, respectively. In the MB approximation,
\begin{align}
    \gamma_{n\to m}^{(S_u)}[a,b,c,d]  =  \frac{1}{S_u}\int  \bigg[\prod_{i=1}^{n+m=5} d\widetilde\Pi_i \bigg] \, (2\pi)^4\delta^{(4)}\left(\sum_{i=1}^n P_i - \sum_{j=n+1}^{n+m=5}P_j\right) (P_a\cdot P_b)(P_c\cdot P_d)\prod_{i=1}^{n}e^{-E_i/T} \,.
\end{align}
Here $d\widetilde\Pi_i= d^3p_i/[(2E_i (2\pi)^3]$, without the factor of $g_i$. We will take into account the degrees of freedom in the end with $\prod_{i}^5 g_i = 2^4 \times 3^3$. The symmetry factor $S_u$ accounts for identical particles, in particular the total number of $u$ quarks or antiquarks in either the initial or final states: $S_u=1$ if there is only one $u$ quark or antiquarks, $S_u=2$ otherwise.

For the $12\to345$ processes
\begin{align}
    \gamma_{uu\to de\phi} &= \frac{1}{18\Lambda^6}\big\{2\gamma_{2\to 3}^{(2)}[1,3,2,4]+2\gamma_{2\to 3}^{(2)}[1,4,2,3] -  \gamma_{2\to 3}^{(2)}[1,2,3,4]\big\}= \frac{4 B_1-B_2}{36 \Lambda^6},\\
        \gamma_{de\to uu\phi} &= \frac{1}{18\Lambda^6}\big\{2\gamma_{2\to 3}^{(2)}[1,3,2,4]+2\gamma_{2\to 3}^{(2)}[1,4,2,3] -  \gamma_{2\to 3}^{(2)}[1,2,3,4]\big\}=\frac{4 B_1-B_2}{36 \Lambda^6}, \\
        \gamma_{ue\to ud\phi} &=\frac{1}{18\Lambda^6}\big\{2\gamma_{2\to 3}^{(1)}[1,3,2,4]+2\gamma_{2\to 3}^{(1)}[1,2,3,4] -  \gamma_{2\to 3}^{(1)}[1,4,2,3]\big\} = \frac{B_1+2B_2}{18\Lambda^6},
        \\
        \gamma_{ude\to ue\phi} &=\frac{1}{18\Lambda^6}\big\{2\gamma_{2\to 3}^{(1)}[1,2,3,4]+2\gamma_{2\to 3}^{(1)}[1,4,2,3] -  \gamma_{2\to 3}^{(1)}[1,3,2,4]\big\} = \frac{B_1+2B_2}{18\Lambda^6}.
\end{align}
Notice that since $1,\, 2$ are initial state particles and $3,\,4$ are always final state particles, due to the symmetry of the rate under relabeling of initial and final states particles, we have $\gamma_{2\to 3}^{(S_u)}[1,3,2,4]=\gamma_{2\to 3}^{(S_u)}[1,4,2,3]\equiv B_1/S_u$. We define instead $\gamma_{2\to 3}^{(S_u)}[1,2,3,4]\equiv B_2/S_u$.  We compute the first building block
\beq 
B_1 =\int \bigg[\prod_{i=1}^{5} d\widetilde\Pi_i \bigg] \, (2\pi)^4\delta^{(4)}\left(P_1 + P_2 -P_3 -P_4-P_5\right)(P_1\cdot P_3)(P_2\cdot P_4)e^{-(E_1+E_2)/T}=\int d\widetilde\Pi_1 d\widetilde\Pi_2 A_1^{(\rm LI)} e^{-(E_1+E_2)/T} \,.
\eeq 
We defined the explicitly Lorentz-invariant component as
\begin{align}
    A_1^{(\rm LI)} \equiv \int\bigg[\prod_{i=3}^{5} d\widetilde\Pi_i \bigg] \, (2\pi)^4\delta^{(4)}\left(P_1 + P_2 -P_3 -P_4-P_5\right)(P_1\cdot P_3)(P_2\cdot P_4) \,.
\end{align}
This allows us to choose a specific frame to perform the calculation of $A_1^{(\rm LI)}$. First, we use the three-dimensional Dirac delta to remove the integration over $\vec p_5$. We choose the frame where $\vec{p}_1 = -\vec{p_2}$, $P_i=\sqrt{s}/2(1,(-1)^{i+1} \hat{p}_1)$ for $i=1,2$ and $P_{j} = E_{j}(1,\hat{p}_{j})$ for $j=3,4$.  $A_1^{(\rm LI)}$ then reads:
\begin{align}
   A_1^{(\rm LI)} =& \frac{\pi s}{2}\int  d\widetilde\Pi_3 d\widetilde\Pi_4  \frac{E_3 E_4}{2E_5}\delta(\sqrt{s}-E_3-E_4-E_5)(1-c_{3;1})(1+c_{4;1})\\
    =&\frac{\pi s}{2}\int d\widetilde\Pi_3 d\widetilde\Pi_4
E_{3}E_{4}\,\,\Theta\left(\sqrt{s}-E_{3}-E_{4}\right)\,\delta\left(\left(\sqrt{s}-E_{3}-E_{4}\right)^{2}-E^{2}_{3}-E^{2}_{4}-2E_{3}E_{4}c_{4;3}\right) \left(1-c_{3;1}\right)\left(1+c_{4;1}\right) \nonumber
\end{align}
where $c_{i;j}$ is the $\cos$ of the angle $\theta_{i;j}$ between $\vec{p}_i$ and $\vec{p}_j$. For the $\sin$ of the angle, we use $s_{i;j}$. Using the identity $c_{4;1}=c_{4;3}c_{1;3}+s_{4;3}s_{1;3}\cos\left(\phi_{4;3}-\phi_{1;3}\right)$, and integrating over the solid angles of $\vec p_{3}$ and $\vec p_{4}$ yields:
\begin{align}
   A_1^{\rm (LI)} =& \frac{s}{16\left(2\pi\right)^{3}}\int   dE_{3}dE_{4}\,\,E_{3}E_{4}\,\,\Theta\left(\sqrt{s}-E_{3}-E_{4}\right)\Theta\left(\left(\sqrt{s}-2E_{3}\right)\left(\sqrt{s}-2E_{4}\right)\right)\nonumber\\
    &\qquad\times\Theta\left( E_{3}+E_{4}-\frac{\sqrt{s}}{2}\right)\left(1-\frac{\left(\sqrt{s}-E_{3}-E_{4}\right)^{2}-E^{2}_{3}-E^{2}_{4}}{6E_{3}E_{4}}\right)   =      \frac{s^{3}}{1024\left(2\pi\right)^{3}}.
    \end{align}
Now we can proceed with the integration to obtain $B_1$
\begin{align}
   B_1 = &  \int d\widetilde\Pi_1 d\widetilde\Pi_2\frac{s^{3}}{1024\left(2\pi\right)^{3}}  e^{-(E_1+E_2)/T}=\frac{1}{2^{13}\left(2\pi\right)^{7}}\int^{\infty}_{0}ds\,\,s^{3}\int^{\infty}_{\sqrt{s}}dE_{+}\int^{\sqrt{E^{2}_{+}-s}}_{-\sqrt{E^{2}_{+}-s}}dE_{-}\exp\left(-\frac{E_{+}}{T}\right)\, = \frac{9T^{10}}{\left(2\pi\right)^{7}}\,,
    & 
\end{align}
where we defined $E_{\pm}\equiv E_1\pm E_2$. The second building block is 
\beq 
B_2 =\int \bigg[\prod_{i=1}^{5} d\widetilde\Pi_i \bigg] \, (2\pi)^4\delta^{(4)}\left(P_1 + P_2 -P_3 -P_4-P_5\right)(P_1\cdot P_2)(P_3\cdot P_4)\,e^{-(E_1+E_2)/T}\,.
\eeq 
To compute it, we proceed as for $B_1$, resulting in
\beq
B_2 = \frac{24T^{10}}{\left(2\pi\right)^{7}}\,.
\eeq
Combining the contributions, the total $2\to 3$ rate is 
\begin{align}
\gamma_{2\to3} = 2^4\times 3^3\times \frac{2B_1 +   B_2}{6 \Lambda^6} = \frac{3024 T^{10}}{(2\pi)^7 \Lambda^6}\,.
\end{align}

For the $123\to45$ processes
\begin{align}
\gamma_{uud\to e\phi} &= \frac{1}{18\Lambda^6}\big\{
2\gamma_{3\to 2}^{(2)}[1,3,2,4]
+2\gamma_{3\to 2}^{(2)}[1,4,2,3] 
-  \gamma_{3\to 2}^{(2)}[1,2,3,4]\big\}= \frac{B_3}{12\Lambda^6}\ ,\\
\gamma_{uue\to d\phi} &=\frac{1}{18\Lambda^6}\big\{
2\gamma_{3\to 2}^{(2)}[1,4,2,3]
+2\gamma_{3\to 2}^{(2)}[1,3,2,4] 
-  \gamma_{3\to 2}^{(2)}[1,2,3,4]\big\}=  \frac{B_3}{12\Lambda^6}\ ,\\
\gamma_{ude\to u\phi} &=\frac{1}{18\Lambda^6}\big\{
2\gamma_{3\to 2}^{(1)}[1,4,2,3]
+2\gamma_{3\to 2}^{(1)}[1,2,3,4] 
-  \gamma_{3\to 2}^{(1)}[1,3,2,4]\big\}= \frac{B_3}{6\Lambda^6}\ .
\end{align}
Here the relabeling symmetry in the rate ensures that swaps of the three initial-state particles do not change the rate. Hence $\gamma_{3\to 2}^{(S_u)}[1,3,2,4]=\gamma_{3\to 2}^{(S_u)}[1,4,2,3]=\gamma_{3\to 2}^{(S_u)}[1,2,3,4]\equiv B_3/S_u$. The last building block is then
\beq 
B_3 =\int \bigg[\prod_{i=1}^{5} d\widetilde\Pi_i \bigg] \, (2\pi)^4\delta^{(4)}\left(P_1 + P_2 +P_3-P_4-P_5\right)(P_1\cdot P_3)(P_2\cdot P_4)\,e^{-(E_1+E_2+E_3)/T}\,.
\eeq 
As we did for $B_1$, we introduce the explicitly Lorentz-invariant part of the integral
\begin{align}
    A_3^{(\rm LI)} \equiv  \int d\widetilde\Pi_4 d\widetilde\Pi_5  (2\pi)^4\delta^{(4)}\left(P_1 + P_2 +P_3-P_4-P_5\right)(P_1\cdot P_3)(P_2\cdot P_4)
\end{align}
so that we can compute it in a frame of choice. Working in the center-of-mass frame of particles $1,2$, we can proceed with the calculation in a way similar to that for $A_1^{(\rm LI)}$, yielding
\begin{align}
   A_3^{\rm (LI)} =& \frac{\sqrt{s_{12}}}{8\left(2\pi\right)}\frac{\left(P_{1}\cdot P_{3}\right)}{E_{3}}\int dE_{4}E_{4}\,\,\Theta\left( E_{3}-E_{4}+\frac{\sqrt{s_{12}}}{2}\right)\Theta\left(E_{4}-\frac{\sqrt{s_{12}}}{2}\right)\nonumber\\
    &\qquad\times\left(1-\frac{\left(\sqrt{s_{12}}+E_{3}-E_{4}\right)^{2}-E^{2}_{3}-E^{2}_{4}}{2E_{3}E_{4}} c_{3;1}\right)   = \frac{P_1\cdot P_3}{16\pi} \left(P_2\cdot P_3 + P_1 \cdot P_2\right),
    \end{align}
where $s_{12}\equiv (P_1 + P_2)^2$. The remaining integrals in $B_3$ are done in a straightforward way using the identity $c_{3;2}=c_{3;1}c_{2;1}+s_{3;1}s_{2;1}\cos\left(\phi_{3;1}-\phi_{2;1}\right)$, yielding 
\beq
B_3 =\int d\widetilde \Pi_1 d\widetilde \Pi_2 d\widetilde \Pi_3 A_3^{(\rm LI)} e^{-(E_1+E_2+E_3)/T} = \frac{6 T^{10}}{\left(2\pi\right)^{7}}\,.
\eeq
Combining the contributions, the total $3\to 2$ rate is 
\begin{align}
\gamma_{3\to2} =  2^4\times 3^3\times \frac{B_3}{3 \Lambda^6} =\frac{864T^{10}}{(2\pi)^7\Lambda^6} . 
\end{align}

The Boltzmann equation is given by
\begin{align}
    \frac{dY_{\phi+\phi^*}}{d\log T} = -\frac{\gamma_\phi^{T\gg T_{\rm PT}}(T)}{s(T) H(T)}\left(1+ \frac{1}{3}\frac{d\log g_{*s}}{d\log T} \right)\,.
\end{align}
Here the total rate is 
\beq 
\gamma_\phi^{T\gg T_{\rm PT}}= 2(\gamma_{2\to 3} +\gamma_{3\to 2}) = \frac{7776T^{10}}{(2\pi)^7\Lambda^6}\,, 
\eeq 
where the factor of $2$ takes into account the production of both $\phi$ and $\phi^*$. 
We can directly integrate the Boltzmann equation to find the relic density
\begin{align}
\frac{m_\phi Y_{\phi+\phi^*}}{0.44\,\rm  eV} & =  \int^{T_{\rm reh}} \frac{dT}{T}\frac{\gamma_\phi^{T\gg T_{\rm PT}}(T)}{H(T) s(T)}\left(1+ \frac{1}{3}\frac{d\log g_{*s}}{d\log T} \right) \simeq \frac{1.4\times10^{23}{\rm GeV} \times T^5_{\rm reh}}{\Lambda^6 }\nonumber 
 \simeq  \left(\frac{y}{1.2\times10^{-21} }\right)^2 \left(\frac{T_{\rm reh}}{\rm 1~\rm TeV }\right)^5\,,\\ 
\end{align}
where we used $y \simeq 0.0144~{\rm GeV}^3 /\Lambda^3\,$ \cite{Aoki:2017puj}. By imposing the quantity above to be about 1 we find $y(T_{\rm reh})$ for the UV freeze-in. Since all these values of $y$ are much smaller than the value required for freeze-in after QCDPT, the UV freeze-in before QCDPT is the dominant freeze-in production channel.

\section{C.~Dark Matter Freeze-out}

The morphology of the freeze-out scenario varies non-trivially depending (i) on the coupling strength $y$ of the effective $pe\phi$ interaction, (ii) the precise hierarchy between the DM mass and the proton mass, and (iii) the initial asymmetry in baryon number. This leads to an interesting interplay between the baryon abundance and the abundance of the complex scalar, despite the couplings being excluded by various constraints.

There are two important types of processes involved in the DM freeze-out: the hadronic channels 
\begin{equation}
    \phi^* +\pi \leftrightarrow N + e^-
\end{equation}
and electromagnetic channels 
\beq \label{fo_EM}
\phi^* + e^+ \leftrightarrow p +\gamma \qquad {\rm and} \qquad \phi^* + \gamma \leftrightarrow p + e^- \,.
\eeq 
The hadronic channels are important at high temperature ($T \gtrsim m_\pi$) due to the large nuclear Yukawa coupling,  but are Boltzmann suppressed for $T\ll m_\pi$. We neglected processes involving more than one pion since they only contribute at the highest temperature after QCDPT (at most) as much as $\phi^* \pi\to Ne^-$ does. The electromagnetic processes are highly effective because electrons and photons remain abundant in the thermal bath down to $T\sim m_e$. Furthermore, the near-degeneracy  $m_\phi \sim m_p$ enhances the low-temperature cross section for $T> m_e$, so that $\left<\sigma v \right> \propto T^{-1} \log T$ and $\Gamma /H  \sim \log T$. Hence, the DM can stay in equilibrium at temperatures much below its mass. 

In order for freeze-out to be viable, the hadronic processes must be efficient enough to maintain equilibrium after the QCDPT, corresponding to couplings $ y \gtrsim 5 \times 10^{-9}$. The hadronic rates will be faster than the electromagnetic rates down to $T\sim 40~{\rm MeV}$, when the pion density becomes too Boltzmann suppressed and the electromagnetic processes take over.  The electromagnetic couplings are only relevant if they can be in equilibrium after the hadronic interactions freeze-out, for $y \gtrsim y_{\rm FO}^{\gamma}=3 \times 10^{-8}$, in which case they stay in equilibrium down to $T=m_e$. 

DM can freeze-out with a mostly symmetric abundance or asymmetric abundance. A strong asymmetric relic abundance is only possible if the DM maintains equilibrium to temperatures $\mathcal{O}(|m_p -m_\phi|) \lesssim m_e$, where the bath will distribute baryon number to the DM and baryons according to the MB distribution. 

\paragraph{Asymmetric freeze-out.} 
For large enough $y$, the electromagnetic couplings will stay in equilibrium until $T \sim m_e$. This is well past proton-antiproton annihilation.

The electromagnetic channels are efficient in converting protons to $\phi^*$ and vice versa, while $\phi$ particles are annihilated away like the antiprotons. As a result, chemical equilibrium between them is assured and the DM closely tracks the baryon abundance as
\beq
\frac{\Omega_{ \phi }}{\Omega_{ p }} \simeq \frac{m_\phi\, n_{ \phi }^{\rm eq}}{m_p \,n_{ p }^{\rm eq}} = \frac{1}{2} \left(\frac{m_\phi}{m_p}\right)^{5/2} e^{-(m_\phi -m_p)/T}\,. \label{focoin}
\eeq
In this case we have \textit{asymmetric freeze-out}. The primordial baryon asymmetry will be split between the baryons and the dark matter. In order to match the observed DM and baryon abundances the asymmetry must be  
\beq
Y_{\phi^*}(T_{\rm ini})+Y_p(T_{\rm ini})-Y_{\phi}(T_{\rm ini})-Y_{\overline p}(T_{\rm ini})=Y_B^{\rm obs }+Y_{\rm DM}^{\rm obs } = 8\times 10^{-11}+ 0.44 {\,\rm eV}/m_\phi
\eeq

Eq.~\eqref{focoin} remains valid as long as the electromagnetic processes are efficient. Eventually, when $T\lesssim m_e$, the mass hierarchy between $\phi^*$ and protons becomes relevant as flagged by the exponential factor in Eq.~\eqref{focoin}. If $m_\phi <m_p$ with $|m_p-m_\phi|\lesssim m_e$, the abundance of $\phi^*$ can receive an enhancement in its abundance relative to baryons that allow it to  reproduce the observed ratio of $\Omega_\phi /\Omega_p \simeq 5.5$. 
Analytically, solving  
\beq 
\langle \sigma v\rangle_{\phi \gamma \to p e}^{T\ll m_e}n_\gamma +\langle \sigma v\rangle_{\phi e \to p \gamma}^{T\ll m_e}  n_e \simeq 2 \langle \sigma v\rangle_{\phi e \to p \gamma}^{T\ll m_e}(T_f) n_e(T_f)\simeq \frac{y^2 \alpha_{\rm EM}(2-\delta)}{m_e m_p}\left(\frac{m_e T_f}{2\pi}\right)^{3/2}e^{-m_e/T_f}= H(T_f)
\eeq 
for $T_f$ and substituting back in Eq.~\eqref{focoin} we get 
\beq 
\frac{\Omega_{ \phi }}{\Omega_{ p }} \simeq \frac{1}{2}\left( \frac{y^2 (2-\delta)}{\alpha_{\rm EM}^{-1}(2\pi)^{3/2} \sqrt{T_f/m_e}\sqrt{\pi^2g_*/90}m_p/M_{\rm Pl}}\right)^{\delta-1}
\simeq 
\frac{1}{2} \left(\frac{y \sqrt{2-\delta}}{2\times 10^{-8}}\right)^{2( \delta-1)}  .
\eeq 

\begin{figure}[t]
    \centering
    \includegraphics[width=0.45\linewidth]{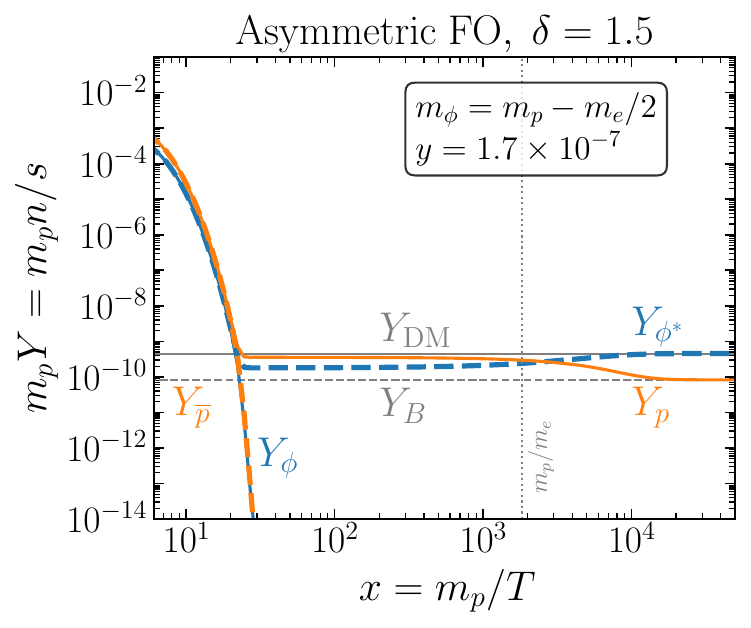} \qquad 
    \includegraphics[width=0.45\linewidth]{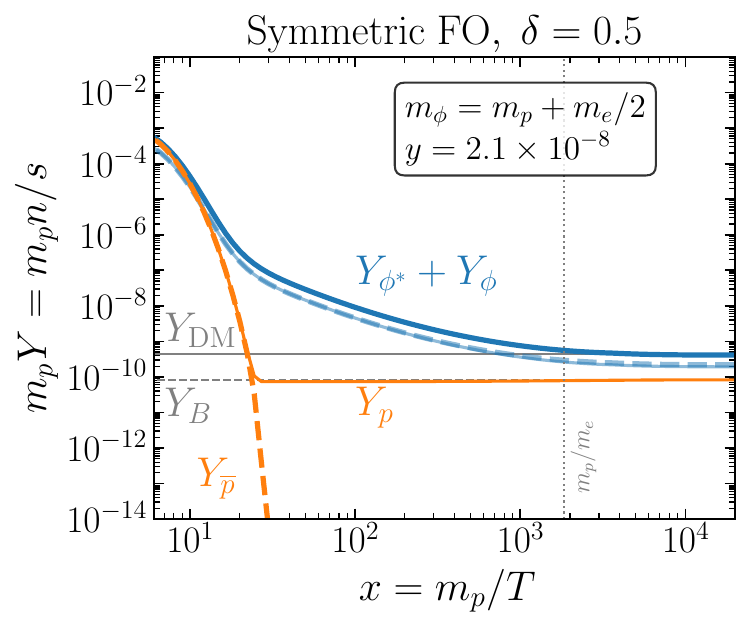}
    \caption{Temperature evolution of DM and proton-antiproton yields in the asymmetric (\textbf{left}) and symmetric (\textbf{right}) freeze-out scenarios. Note the coupling to reproduce the relic density differs by one order of magnitude between the two cases, even if the DM mass only varies by $\sim m_e$ (less then one per mille).}
    \label{fig:FO}
\end{figure}

The freeze-out for a given choice of $m_\phi$ and $y$ is shown in the left panel of Fig.~\ref{fig:FO}. Interestingly, asymmetric freeze-out would allow to avoid direct detection constraints: the DM abundance is entirely comprised of $\phi^*$ and only antimatter targets would give a possibility for direct detection.

\paragraph{Symmetric freeze-out.} 
After the hadronic processes become inefficient and protons and antiprotons annihilate, the $\phi$ and $\phi^*$ can keep on annihilating by converting into baryons via the electromagnetic reactions~\eqref{fo_EM}. We dub this scenario \textit{symmetric freeze-out} as the DM density will be given by $Y_{\phi}+Y_{\phi^*}\simeq 2 Y_{\phi}$. The annihilation of $\phi,\phi^*$ happens slowly, as evident in the right panel of Fig.~\ref{fig:FO}. As the temperature drops, conversion of $\phi,\phi^*$ into baryons and antibaryons become  inefficient and the leftover baryons cannot produce additional DM if inverse reactions are quenched by choosing $m_\phi>m_p$. This allows both the baryons and the DM to reproduce the relic abundances. Note that a small asymmetry of about 20\% between the abundances of $\phi$ and $\phi^*$ is still present and has to be inserted into the initial conditions in order to match the observed baryon density. The asymmetry produces a considerable fraction of protons at temperatures below the electron mass.

\section{D.~Dark Matter Production in Neutron Stars}
In this section, we characterize the neutron star (NS) environment and calculate the rate of DM production and the resulting energy injection in the core of these extremely dense objects. 

\subsection{Neutron star environment}
 \begin{figure}[t]
     \centering
     \includegraphics[width=0.9\linewidth]{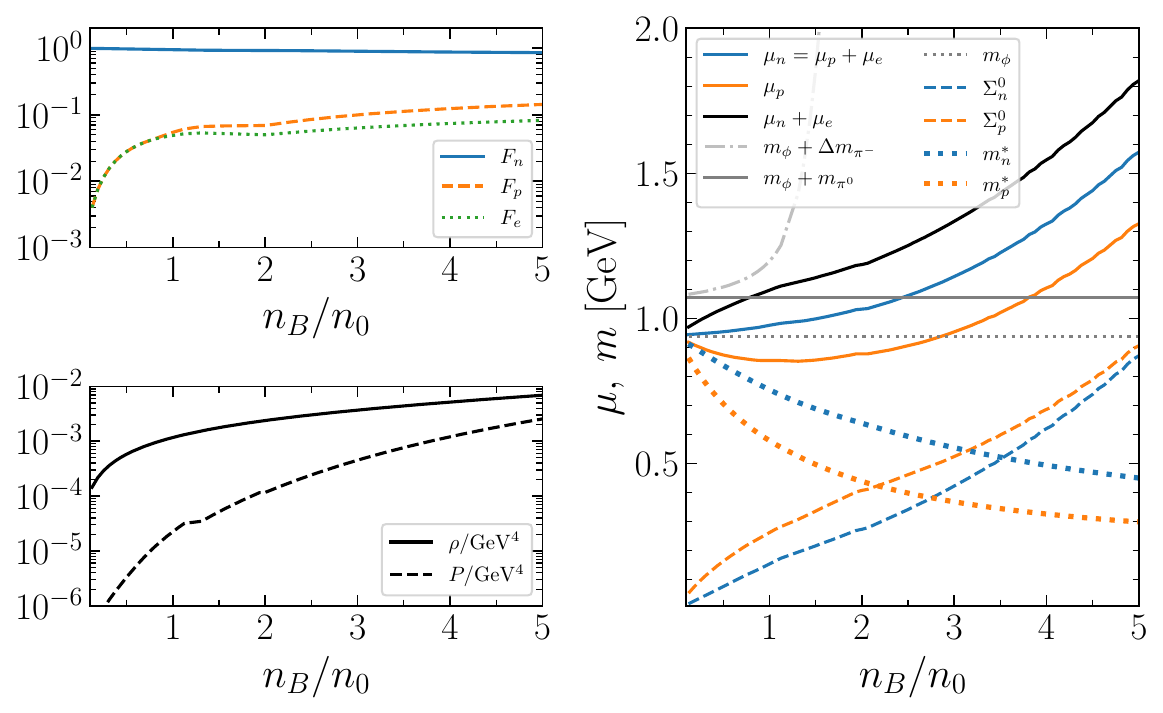}
     \caption{NS properties extracted from the APR EOS \cite{Akmal:1998cf} provided by NSCool \cite{NSCool} and their dependence on the baryon number density $n_B$ normalized to the saturation value $n_0=0.16\ {\rm fm^{-3}}$. \textbf{Left}: the quantities reported in the data file that define the properties of the NS core: densities of various particle species $F_i$ in units of $n_B$ (upper plot) and energy density $\rho$ and pressure $P$ in ${\rm GeV}^4$ (lower plot). \textbf{Right}: the extracted chemical potentials $\mu_i$, effective masses $m_i^*$ and self energies $\Sigma^0_i$ and comparison to mass thresholds. Note that due to the uncertainty in the mass of the charged pion~\cite{Fore:2023gwv}, $\mu_n+\mu_e <m_\phi +\Delta m_{\pi^-}$, we conservatively neglect the processes involving charged pions. The channel \eqref{pion0_process} is active only when $\mu_n=\mu_p+\mu_e > m_\phi+m_{\pi^0}$, which happens for $n_B\gtrsim2.5\,n_0$. Instead, the process \eqref{gamma_process} is active for all the values of $n_B$ we consider here. }
     \label{fig:NS}
 \end{figure}

The chemical potentials and the densities of the NS species can be obtained from an equation of state, such as the Akmal, Pandharipande, and Ravenhall (APR) model \cite{Akmal:1998cf}. We use the APR equation of state data together with the catalyzed crust model file provided in the \texttt{NSCool} code \cite{NSCool}, where all the quantities depend on the total baryon density $n_B$; we show this data in the left plot of Fig.~\ref{fig:NS}. For density $\rho(n_B)$ and pressure $P(n_B)$, the total baryon chemical potential is obtained via (see e.g. \cite{Brandes:2023bob})
\beq 
\mu = \frac{\partial \rho}{\partial n_B} \simeq \frac{\rho +P}{n_B}, 
\eeq 
where the last equality is valid at zero temperature. Since $\mu = F_n \mu_n+F_p\mu_p$, where $F_i=n_i/n_B$ are the relative densities and $\mu_n= \mu_p+\mu_e$, we can obtain the chemical potentials as 
\begin{subequations}
    \begin{align}
    \mu_e =& \sqrt{m_e^2+{(p_e^{\rm F})}^2}=\sqrt{m_e^2 + (3\pi^2 n_B F_e)^{2/3}}\,,\\
       % \mu_\mu =& \sqrt{m_\mu^2+{(p_\mu^{\rm F})}^2}=\sqrt{m_\mu^2 + (3\pi^2 n_B F_\mu)^{2/3}}\quad {\rm if}\,{\mu_e>m_\mu}\,,\\
        \mu_n =& \frac{\mu+F_p\mu_e}{F_n+F_p}\,,\\
        \mu_p =& \frac{\mu-F_n\mu_e}{F_n+F_p} \,.
    \end{align}
\end{subequations}
We show the calculated chemical potentials in the right plot of Fig.~\ref{fig:NS} and compare them to other quantities: $m_\phi\simeq m_p$ and $m_\phi+m_{\pi^0}$ mass thresholds, the upper bound on the final state mass threshold $m_\phi+\Delta m_{\pi^-}$ and self-energy of nucleons in the NS frame $\Sigma^0_{n} $, $\Sigma^0_{p} $ calculated from the effective masses given in the APR equation of state, ${m_p^*}$ and $m_n^*$. The dispersion relation of baryons can be assumed to be~\cite{Berryman:2023rmh}
\beq 
E_B(p_B) =  \sqrt{{m_B^*(n_B)}^2 + |\vec{p}_B|^2} + \Sigma^0(n_B)\, ;\eeq 
the self-energies $\Sigma^0(n_B)$ can be obtained by enforcing $E_B(p_B^{\rm F})=\mu_B(n_B)$, with $p_B^{\rm F}=  (3\pi^2 n_B F_B)^{1/3}$. Due to in-medium effects, the equations of motion for the nucleon fields get modified \cite{Berryman:2023rmh}. One can show that these equations of motion can be written in the form of free Dirac equations, provided that the canonical four-momenta $p^\mu$ of the baryons are replaced by in-medium momenta $p^\mu \to p^*_\mu = p_\mu - \Sigma_\mu$ and vacuum masses are replaced with effective masses $m_B\to m_B^*$. We ignore frame effects due to the rotation of the star: the self-four-momentum is taken as $\Sigma_\mu=(\Sigma_0,\pmb \Sigma)=(\Sigma_0,\pmb 0)$. Due to the different normalization of the Dirac fields, the Lorentz-invariant phase space for baryons involves $E^*_B=p^*_0$ instead of $E_B$ in the denominator: $d\Pi_B^* = g_id^3p/(2E^*_B)/(2\pi)^3$.  

\subsection{Processes}
The $\phi p e$ vertex allows for the following proton conversion processes in the core of a NS:
\begin{subequations}
\begin{align}
    \label{gamma_process}
   p + e^-\to&\ \gamma +\phi^* \,,\\
    \label{pion0_process}
    p + e^-\to&\ \pi^0+\phi^* \,,\\
    \label{npe_process}
   n+p + e^-\to&\ n +\phi^* \,,\\
    \label{ppe_process}
   p+p + e^-\to&\ p +\phi^* \,,\\
   \label{np_process}
   n+p \to&\ e^++ n +\phi^* \,,\\
   \label{pp_process}
   p+p \to&\  e^++ p +\phi^* \,.
\end{align}
\end{subequations}
Binary scatterings can produce photons or neutral pions in the final state, depositing energy in the neutron star. $\phi^*$ can also be produced via a bystander neutron or proton: these have either an electron in the initial state or a positron in the final state, that can later deposit energy in the NS via annihilation with an electron.
Conservatively, we neglected processes involving charged pions in the initial or final state, due to the huge uncertainty of the charged pion dispersion relations in a nuclear density medium~\cite{Fore:2023gwv}, which can prevent the process from happening in most scenarios. If these processes were relevant they would make $\phi^*$ production even more efficient.  Since $\beta$-equilibrium is enforced ($\mu_n=\mu_p+\mu_e$), proton conversion into $\phi^*$ ultimately leads to a neutron replacing the proton and hence to neutron conversion.

The interactions among nucleons in the inner core of a NS cannot be modeled by the simple pion-nucleon chiral Lagrangian Eq.~\eqref{IRlagrangian}.  To describe the interactions between nucleons in the NS we follow the standard approach in the literature of considering a nuclear potential (see e.g. \cite{Friman:1979ecl,Bottaro:2024ugp,Fiorillo:2025zzx}). To retain the phenomenological properties of the Fermi liquid approach but not restricting to the fully non-relativistic case, since our effective vertex $\phi p e$ contains a relativistic electron, we consider the following Lagrangian
\begin{align}
{\cal L}_{NNNN} = &- \frac{C_f}{2}(\overline N N)(\overline N N) - \frac{C_{f'}}{2}(\overline N  {\pmb\tau} N)\cdot(\overline N  {\pmb\tau} N)\\\nonumber
&+ \frac{C_{g}}{2}(\overline N \gamma^\mu
\gamma^5 N)(\overline N 
\gamma_\mu \gamma^5 N) + \frac{C_{g'}}{2}(\overline N \gamma_\mu \gamma^5{\pmb\tau} N)\cdot(\overline N \gamma^\mu \gamma^5 {\pmb\tau} N) + \frac{C_{h'}}{2}(\overline N   \gamma^5{\pmb\tau} N)\cdot(\overline N   \gamma^5 {\pmb\tau} N)\ ,
\end{align}
where $N=(p,n)^T$ is the isospin doublet and ${\pmb\tau}$ are the isospin Pauli matrices. In the non-relativistic limit, this Lagrangian matches the following nuclear potential
\begin{align}
    V_{N_1N_2}({\bf k}) = f + f' {\pmb\tau}_1 \cdot {\pmb\tau}_2 
    + g {\pmb\sigma}_1\cdot {\pmb  \sigma}_2 
    + g'_k ({\pmb\tau}_1 \cdot {\pmb\tau}_2) ({\pmb\sigma}_1\cdot {\pmb  \sigma}_2) + h'_k ({\pmb \sigma_1}\cdot \widehat{\bf k} )({\pmb \sigma_2}\cdot \widehat{\bf k} ) ({\pmb\tau}_1 \cdot {\pmb\tau}_2),
\end{align}
where $\pmb \sigma$ are the spin operator Pauli matrices, and the couplings dependent on momentum-transfer read
\begin{align}
g_k' &= g' -C_\rho \frac{f_{\pi NN}^2}{m_\pi^2}  \frac{k^2}{k^2+m_\rho^2},\\
h_k' &= -\frac{f_{\pi NN}^2}{m_\pi^2}\left(\frac{k^2}{k^2+m_\pi^2}-C_\rho \frac{k^2}{k^2+m_\rho^2}\right).
\end{align}
We take $C_\rho =  1.4$, $f_{\pi NN} =  g_A m_\pi / (2F_\pi) =0.96$ and $m_\rho =  770$ MeV. The Landau parameters are given as \cite{Bottaro:2024ugp,Fiorillo:2025zzx}
$$\{f,f',g,g'\} = \frac{\pi^2}{ 2m_n^* p^{\rm F}_n}\{F_0,F_0',G_0,G_0'\},$$
with $F_0=0.5$ $F_0'=0.7$, $G_0=G_0'=1.1$.
The Lagrangian parameters match to the nuclear potential parameters as 
\begin{align}
    \{C_f,C_{f'},C_g,C_{g'}(k), C_{h'}(k)  \} = \{ f,f',g, g'_{k}, \frac{4{m_n^*}^2}{k^2}h'_{k}\}.
\end{align}
Note that the isospin bilinears are expanded as
\begin{align}
& (\overline N  \Gamma  N)(\overline N \Gamma   N)= (\overline n\Gamma n)^2 
    +
    (\overline p\Gamma p)^2
    +
   2 (\overline n\Gamma n)(\overline p\Gamma p),\\
   & (\overline N  \Gamma {\pmb\tau} N)(\overline N \Gamma {\pmb\tau} N) = (\overline n\Gamma n)^2 
    +
    (\overline p\Gamma p)^2
    -
   2 (\overline n\Gamma n)(\overline p\Gamma p) + 4 (\overline n\Gamma p)(\overline p\Gamma n) .
\end{align}
This allows us to write the Feynman rules for the vertices of interest. The four-nucleon vertices are 
 \begin{align} \overline p_{s_1}(k_1)p_{s_2}(k_2)\overline p_{s_3}(k_3)p_{s_4}(k_4)\longrightarrow {i}\Big[&-(C_f+C_{f'})(\delta_{s_1s_2} \delta_{s_3s_4}- \delta_{s_1s_4} \delta_{s_2s_3})\\\nonumber
&+(C_g+C_{g'}(k_1-k_2))(\gamma^\mu \gamma^5)_{s_1s_2}(\gamma_\mu \gamma^5)_{s_3s_4}\\\nonumber
&-(C_g+C_{g'}(k_1-k_4))(\gamma^\mu \gamma^5)_{s_1s_4}(\gamma_\mu \gamma^5)_{s_2s_3}\\\nonumber
&+C_{h'}(k_1-k_2) \gamma^5_{s_1s_2}\gamma^5_{s_3s_4}- C_{h'}(k_1-k_4)\gamma^5_{s_1s_4}\gamma^5_{s_2s_3}
\Big]
\end{align}

\begin{align} \overline n_{s_1}(k_1)n_{s_2}(k_2)\overline p_{s_3}(k_3)p_{s_4}(k_4)\longrightarrow i\Big[-&(C_f-C_{f'})\delta_{s_1 s_2}\delta_{s_3s_4}+2 C_{f'}\delta_{s_1s_4}\delta_{s_2s_3}\\\nonumber
    +&(C_g-C_{g'}(k_1-k_2))(\gamma^\mu \gamma^5)_{s_1s_2}(\gamma_\mu \gamma^5)_{s_3s_4}\\\nonumber
    -&2C_{g'}(k_1-k_4)(\gamma^\mu \gamma^5)_{s_1s_4}(\gamma_\mu \gamma^5)_{s_2s_3}\\\nonumber
    -&C_{h'}(k_1-k_2)\gamma^5_{s_1s_2}\gamma^5_{s_3s_4}-2C_{h'}(k_1-k_4)\gamma^5_{s_1s_4}\gamma^5_{s_2s_3}\Big]\end{align}
    The trilinear vertices are obtained from the Lagrangian~\eqref{IRlagrangian}:
    \begin{align}\overline p_{s_1} p_{s_2}\pi^0(p_3) &\longrightarrow \frac{g_A}{2F_\pi}(\gamma_\mu\gamma^5)_{s_1 s_2}{p}_3^\mu\\
    % \overline p_{s_1} n_{s_2}\pi^+(p_3) &\longrightarrow \frac{g_A}{\sqrt 2F_\pi}(\slashed{p}_3\gamma^5)_{s_1 s_2}\\
    \overline e_{s_1}^c p_{s_2}\phi &\longrightarrow i y (P_R)_{s_1s_2}\\
    \overline e_{s_1} e_{s_2} A^\mu  &\longrightarrow -ie\gamma^\mu_{s_1 s_2}
    \\
    \overline p_{s_1} p_{s_2} A^\mu  &\longrightarrow ie\gamma^\mu_{s_1 s_2}
    \end{align}

Using the above Feynman rules we compute the matrix elements for the processes \eqref{gamma_process}--\eqref{pp_process}. We explicitly write the mass and momenta that enter the various spinors to account for the in-medium modifications. In what follows, we employ the notation
$S_f(p,m)=i[p_\mu \gamma^\mu + m]/(p^2-m^2)$
for a fermion propagator.
\begin{itemize}
    \item $p(p_1)+e^-(p_2) \to \gamma(p_3)+\phi^*(p_4)$. We write
\begin{align}
    |{\cal M}_{pe\to \gamma\phi}|^2 =\frac{1}{g_e g_pg_\gamma g_\phi} \sum_{\rm pols}|{\cal M}_{p}+{\cal M}_{e}|^2. 
\end{align}
where
\begin{align}
    {\cal M}_{p} &= {-ey} \epsilon_\nu(p_3) \overline{v}(p_2,m_e)P_RS_f(p_1^*-p_3,m_p^*)\gamma^\nu u(p_1^*,m_p^*),\\
    {\cal M}_{e} &= { ey} \epsilon_\nu(p_3) \overline{v}(p_2,m_e)P_RS_f(p_2-p_3,m_e)\gamma^\nu u(p_1^*,m_p^*).
\end{align}
\item $p(p_1)+e^-(p_2) \to \pi^0(p_3)+\phi^*(p_4)$. We write
\begin{align}
    |{\cal M}_{pe\to \pi\phi}|^2 =\frac{1}{g_e g_pg_\pi g_\phi} \sum_{\rm pols}|{\cal M}_{\pi}|^2,
\end{align}
where
\begin{align}
  {\cal M}_{\pi} =  \frac{ig_A y}{2F_\pi}  \overline{v}(p_2,m_e)P_R S_f(p_1^*-p_3,m_p^*)\gamma_\mu p_3^\mu  \gamma^5 u(p_1^*,m_p^*) .
\end{align}
\item $n(p_1)+p(p_2)+e^-(p_3) \to n(p_4)+\phi^*(p_5)$. We write
\begin{align}
    |{\cal M}_{npe\to n\phi}|^2 =\frac{1}{g_n^2 g_p  g_e g_\phi} \sum_{\rm pols}|{\cal M}_{nnpp}+{\cal M}_{npnp}|^2,
\end{align}
where 
\begin{align}
    {\cal M}_{nnpp} = iy\bigg[& -(C_f-C_{f'})[\overline u(p_4^*,m_n^*) u(p_1^*,m_n^*)][\overline v(p_3,m_e)P_RS_f(p_1^*+p_2^*-p_4^*,m_p^*)u(p_2^*,m_p^*)]\\\nonumber
    &+(C_g-C_{g'}(|{\bf p}_1-{\bf p}_4|))[\overline u(p_4^*,m_n^*) \gamma^\mu \gamma^5u(p_1^*,m_n^*)][\overline v(p_3,m_e)P_RS_f(p_1^*+p_2^*-p_4^*,m_p^*)\gamma_\mu \gamma^5u(p_2^*,m_p^*)]\\\nonumber
    &-C_{h'}(|{\bf p}_1-{\bf p}_4|)[\overline u(p_4^*,m_n^*)   \gamma^5u(p_1^*,m_n^*)][\overline v(p_3,m_e)P_RS_f(p_1^*+p_2^*-p_4^*,m_p^*) \gamma^5u(p_2^*,m_p^*)]
    \bigg],
\end{align}
\begin{align}
    {\cal M}_{npnp} = 2iy\bigg[& C_{f'}[\overline u(p_4^*,m_n^*) u(p_2^*,m_p^*)][\overline v(p_3,m_e)P_RS_f(p_1^*+p_2^*-p_4^*,m_p^*)u(p_1^*,m_n^*)]\\\nonumber
    &-C_{g'}(|{\bf p}_2-{\bf p}_4|)[\overline u(p_4^*,m_n^*) \gamma^\mu \gamma^5u(p_2^*,m_p^*)][\overline v(p_3,m_e)P_RS_f(p_1^*+p_2^*-p_4^*,m_p^*)\gamma_\mu \gamma^5u(p_1^*,m_n^*)]\\\nonumber
    &-C_{h'}(|{\bf p}_2-{\bf p}_4|)[\overline u(p_4^*,m_n^*)   \gamma^5u(p_2^*,m_p^*)][\overline v(p_3,m_e)P_RS_f(p_1^*+p_2^*-p_4^*,m_p^*) \gamma^5u(p_1^*,m_n^*)]
    \bigg].
\end{align}
\item $p(p_1)+p(p_2)+e^-(p_3) \to p(p_4)+\phi^*(p_5)$. We write
\begin{align}
    |{\cal M}_{ppe\to p\phi}|^2 =\frac{1}{g_p^3 g_e g_\phi} \sum_{\rm pols}|{\cal M}_{14}+{\cal M}_{24}|^2 ,
\end{align}
where
\begin{align}
    {\cal M}_{14} = iy\bigg[& -(C_f+C_{f'})[\overline u(p_4^*,m_p^*) u(p_1^*,m_p^*)][\overline v(p_3,m_e)P_RS_f(p_1^*+p_2^*-p_4^*,m_p^*)u(p_2^*,m_p^*)]\\\nonumber
    &+(C_g+C_{g'}(|{\bf p}_1-{\bf p}_4|))[\overline u(p_4^*,m_p^*) \gamma^\mu \gamma^5u(p_1^*,m_p^*)][\overline v(p_3,m_e)P_RS_f(p_1^*+p_2^*-p_4^*,m_p^*)\gamma_\mu \gamma^5u(p_2^*,m_p^*)]\\\nonumber
    &+C_{h'}(|{\bf p}_1-{\bf p}_4|)[\overline u(p_4^*,m_p^*)   \gamma^5u(p_1^*,m_p^*)][\overline v(p_3,m_e)P_RS_f(p_1^*+p_2^*-p_4^*,m_p^*) \gamma^5u(p_2^*,m_p^*)]
    \bigg],
\end{align}
\begin{align}
  {\cal M}_{24}=   iy\bigg[& (C_f+C_{f'})[\overline u(p_4^*,m_p^*) u(p_2^*,m_p^*)][\overline v(p_3,m_e)P_RS_f(p_1^*+p_2^*-p_4^*,m_p^*)u(p_1^*,m_p^*)]\\\nonumber
    &-(C_g+C_{g'}(|{\bf p}_2-{\bf p}_4|))[\overline u(p_4^*,m_p^*) \gamma^\mu \gamma^5u(p_2^*,m_p^*)][\overline v(p_3,m_e)P_RS_f(p_1^*+p_2^*-p_4^*,m_p^*)\gamma_\mu \gamma^5u(p_1^*,m_p^*)]\\\nonumber
    &-C_{h'}(|{\bf p}_2-{\bf p}_4|)[\overline u(p_4^*,m_p^*)   \gamma^5u(p_2^*,m_p^*)][\overline v(p_3,m_e)P_RS_f(p_1^*+p_2^*-p_4^*,m_p^*) \gamma^5u(p_1^*,m_p^*)]
    \bigg] .
\end{align}
\item $n(p_1)+p(p_2)\to e^+(p_3)  +n(p_4)+\phi^*(p_5)$. We can obtain the matrix element directly from crossing symmetry:
\begin{align}
    |{\cal M}_{np\to en\phi}|^2 =  -|{\cal M}_{npe\to n\phi}|^2(p_3\to -p_3)\ .
\end{align}
\item $p(p_1)+p(p_2)\to e^+(p_3)  +p(p_4)+\phi^*(p_5)$. Also this matrix element can be obtained from crossing symmetry:
\begin{align}
    |{\cal M}_{pp\to ep\phi}|^2 =  -|{\cal M}_{ppe\to p\phi}|^2(p_3\to -p_3).
\end{align}
\end{itemize}

\subsection{Proton conversion rates}
We are interested in collision integrals of the type
\begin{align}\label{eq:general_integral}
    C_{m\to n}[{\cal F}]= \int \left[ \prod_{i=1}^{n+m} d\Pi_i^*\right] (2\pi)^4 \delta^{(4)}\left(\sum_{i=1}^m P_i -\sum_{j=m+1}^{m+n} P_j \right) {\cal F}(P_i^*\cdot P_j^*,m_i^*,E_i^*,\mu_i,T)\ .
\end{align}
Here ${\cal F}$ is any function of four-momentum scalar products, masses, energies in the NS frame, chemical potentials and temperature. The ${^*}$ applies if a particle $i$ is a nucleon: if not, we can simply take $\Sigma_i^0=0$ and $m_i^*=m_i$ so that $E_i^*=E_i$ follows. Note that in general the canonical four-momenta squared have modified dispersion relations $P_i^2=p^\mu_i p_\mu^i = {E_i}^2-p_i^2={m_i^*}^2+ {\Sigma_0^i}^2+2{E_i^*}\Sigma_0^i$. We show in Section~E how to compute integrals like in Eq.~\eqref{eq:general_integral} using Monte Carlo techniques. 

In our notation, the proton conversion rate for the process $\rm P$ are obtained as 
\beq 
\gamma_{\rm P}=C_{m\to n}[|{\cal M}_{\rm P}|^2f_1\dots f_m (1\pm f_{m+1})\dots(1\pm f_{m+n})],\eeq
i.e. the collision integral is evaluated over the squared matrix element times the phase space distribution of initial state particles and Pauli-blocking (or Bose-enhancements, depending on the statistics) for the final state particles.

Explicitly, the binary scatterings are
\begin{align}
    \gamma_{pe \to \gamma \phi} &= \int d\Pi^*_p d\Pi_e d\Pi_\gamma d\Pi_\phi (2\pi)^4 \delta^{(4)}(P_p+P_e-P_\gamma -P_\phi)|{\cal M}_{pe\to \gamma \phi}|^2 f_p f_e,\\
    \gamma_{pe \to \pi \phi} &= \int d\Pi^*_p d\Pi_e d\Pi_\pi d\Pi_\phi (2\pi)^4 \delta^{(4)}(P_p+P_e-P_\pi -P_\phi)|{\cal M}_{pe\to \pi \phi}|^2 f_p f_e,
\end{align}
the proton conversion rates with bystander nucleons are
\begin{align}
    \gamma_{npe \to n \phi} &= \int d\Pi^*_n d\Pi_p^* d\Pi_e d{\Pi_n^*}' d\Pi_\phi (2\pi)^4 \delta^{(4)}(P_n+P_p+P_e-P_n' -P_\phi)|{\cal M}_{npe\to n \phi}|^2 f_n(E_n) f_p(E_p) f_e(E_e) [1-f_n(E_n')],\\
    \gamma_{ppe \to p \phi} &= \frac{1}{2}\int d\Pi^*_p {d\Pi_p^*}' d\Pi_e d{\Pi_p^*}'' d\Pi_\phi (2\pi)^4 \delta^{(4)}(P_p+P_p'+P_e-P_p'' -P_\phi)|{\cal M}_{ppe\to p \phi}|^2 f_p(E_p) f_p(E_p') f_e(E_e) [1-f_p(E_p'')],\\
    \gamma_{np \to e n \phi} &= \int d\Pi^*_n d\Pi_p^* d\Pi_e d{\Pi_n^*}' d\Pi_\phi (2\pi)^4 \delta^{(4)}(P_n+P_p-P_e-P_n' -P_\phi)|{\cal M}_{np\to e n \phi}|^2 f_n(E_n) f_p(E_p)   [1-f_n(E_n')]\\
    \gamma_{pp \to ep \phi} &=\frac{1}{2} \int d\Pi^*_p {d\Pi_p^*}' d\Pi_e d{\Pi_p^*}'' d\Pi_\phi (2\pi)^4 \delta^{(4)}(P_p+P_p'-P_e-P_p'' -P_\phi)|{\cal M}_{pp\to ep \phi}|^2 f_p(E_p) f_p(E_p')  [1-f_p(E_p'')]
\end{align}
The processes with two protons in the initial state are suppressed by a symmetry factor of $1/2$. In all the scatterings above, the $f_i$ indicate Fermi-Dirac distributions for neutrons, protons and electrons in the NS at temperature $T$. Specifically, for $i=\{n,p,e\}$
\begin{align}
    &f_i(E_i) = \frac{1}{e^{(E_i-\mu_i)/T}+1}\ .
\end{align}
Since for old neutron stars $T\ll \mu_e \simeq 200$ MeV, we can take the $T\to 0$ limit, and therefore approximate $f_i(E_i) \simeq \Theta(\mu_i-E_i)$. We neglected the Bose-enhancement factors in all the processes, due to the low phase space occupations of the final state particles. The squared amplitudes of the processes are computed analytically using the Feynman rules shown above. In most cases the expressions are lengthy so we do not show them here.

\begin{figure}
    \centering
\includegraphics[width=\linewidth]{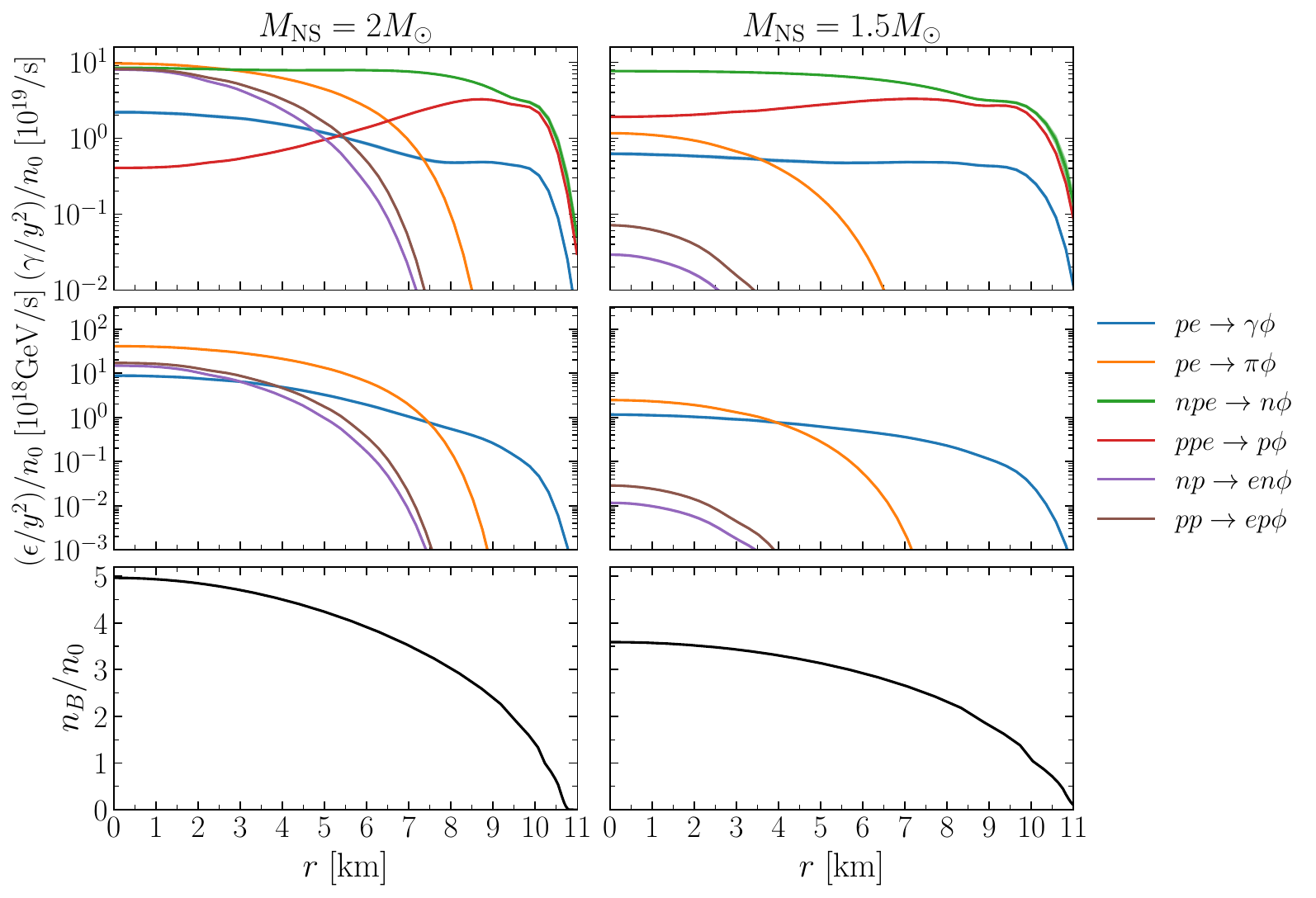}
    \caption{Radial dependence of the proton conversion rates (\textbf{top}), energy injection rates (\textbf{center}) and baryon number density profile in units of saturation density $n_0=0.16\ {\rm fm^{-3}}$ (\textbf{bottom}). We consider two neutron stars with different mass of $2\,M_\odot$ (left plots) and  $1.5\, M_\odot$ (right plots) with density profiles obtained from the APR EOS with catalyzed crust model from \texttt{NSCool}~\cite{NSCool}.  }
    \label{fig:rates_nB_r}
\end{figure}

We show in the top panels of Figure~\ref{fig:rates_nB_r} the radial dependence of the rates for all the considered processes. We distinguish the cases of $2M_\odot$ and $1.5M_\odot$ neutron stars, with density profiles shown in the bottom panels.

Before proceeding, it is important to establish the fate of the $\phi^*$ particles produced via the proton conversion processes. To do that, we compute the fraction $F_{v_\phi>v_{\rm esc}}$ of particles that, in the frame where protons have energy $E_p$, have velocity $v_\phi(E_\phi) = \sqrt{1-m_\phi^2/E_\phi^2}$ larger then the NS escape velocity $v_e=\sqrt{GM_{\rm NS}/R_{\rm NS}}$. We therefore calculate for each process $\rm P$ the ratio
\begin{align}
    F_{v_\phi>v_{\rm esc}}^{\rm P} = \frac{1}{\gamma_{\rm P}}   C_{m\to n}[|{\cal M}_{\rm P}|^2f_1\dots f_m (1\pm f_{m+1})\dots(1\pm f_{m+n})\Theta(v_\phi(E_\phi) - v_{\rm esc})] \,,
\end{align}
The Heaviside theta $\Theta(v_\phi(E_\phi) - v_{\rm esc})$ takes care of selecting the particles that will escape the NS. This fraction is strongly dependent on the NS mass and radius entering the escape velocity. For the two stars with density profiles in Fig.~\ref{fig:rates_nB_r}, the escape velocity is very high: $v_{\rm esc}(2M_{\odot})\simeq 0.73$ and $v_{\rm esc}(1.5M_{\odot})\simeq 0.62$. After the numerical evaluation, we obtain that only less than $3\%$ of $\phi^*$ particles produced in the $2M_\odot$ star escape, mostly produced via the $npe\to n\phi$ and $ppe\to p\phi$ processes. For the $1.5M_\odot$ NS case, the escape fraction is even smaller ($<0.2\%$). We conclude that in both stars most of the $\phi^*$ get trapped in the NS gravitational potential, allowing them, if produced efficiently, to form a condensate. In other words, the mass of the NS does not change much, but the Fermi degeneracy pressure that sustains the star against gravity is reduced.

Given that $\phi^*$ particles are trapped, the star composition is changed drastically if proton conversions are efficient. To put a constraint on the interaction coupling $y$, we require that only $10\%$ of neutrons gets converted throughout the NS lifetime. If this were the case, it would be impossible for the NS to sustain a mass of $>2 M_{\odot}$ \cite{McKeen:2018xwc} until today, unless self-interactions among $\phi^*$ particles are introduced. We can compute the number of converted neutrons and total number of neutrons by integrating over the radial density profile of a NS. We obtain such a profile from the APR EOS with the catalyzed crust model for two stars of mass $2\, M_\odot$ and $1.5\, M_\odot$ (see the bottom panel of Figure~\ref{fig:rates_nB_r}). We compute the total conversion rate as 
\beq 
\gamma_{\rm tot} = \gamma_{pe\to \gamma \phi} + \gamma_{pe\to \pi\phi}+ \gamma_{npe\to n\phi}+\gamma_{ppe\to p\phi}+ \gamma_{np\to e n\phi}+\gamma_{pp\to ep\phi}
\eeq 
and we require $N_{n}^{\rm conv}(\tau_{\rm NS})<0.1 N_n$ with
\begin{align}
N_{n}^{\rm conv}= \tau_{\rm NS}\int dr \ 4\pi r^2\gamma_{\rm tot}(n_B(r)) \qquad {\rm and} \qquad 
N_n=\int dr \ 4\pi r^2 n_n(r)\ ,
\end{align} 
where $\tau_{\rm NS}$ is the age of the NS. We approximated our conversion rates as time-independent since they are computed in the $T\to 0$ limit.
The computed rates over the NS density profile are shown in the top panel of Fig.~\ref{fig:rates_nB_r}. We consider an old NS like PSR J0740+6620~\cite{NANOGrav:2019jur,Fonseca:2021wxt,Riley:2021pdl,Brandes:2023hma} of mass $\sim2\,M_\odot$ and age $\tau_{\rm NS}\simeq 5$ Gyr. Imposing $N_{n}^{\rm conv}<0.1 N_n$ for this star gives a bound of \beq y< 1.2 \times 10^{-19}.\eeq

Consider now a star modeling the coldest known neutron star PSR J2144--3933, having a mass $\sim1.5 \,M_\odot$ and estimated age of $\tau_{\rm NS} \simeq 0.3$ Gyr \cite{Guillot:2019ugf}.
Imposing $N_{n}^{\rm conv}<0.1 N_n$ gives an estimate of the values of the coupling where the modification of the neutron star EOS is perturbative. In this case, we obtain $y<6\times 10^{-19}$.

\subsection{Neutron star heating}
The total energy balance requires that the NS cools down,
%%%%%%%%%%%%%%%%%%%%%%%%%%%%%%
\begin{equation}
    {\cal C} \frac{dT_c}{dt}=-L_\gamma-L_\nu+H\, ;
\end{equation}
where the photon luminosity is related to the surface temperature $T_{\rm NS}$ as $L_{\gamma}\simeq 4\pi \sigma_{\rm SB} R_{\rm NS}^2 T_{\rm NS}^4$, $\sigma_{\rm SB}=\pi^2/60$ is the Stefan-Boltzmann constant, $t$ is the time, $\cal C$ and $T_c$ are the heat capacity and temperature of the NS interior, respectively. All quantities are measured by a distant observer. The final term $H$ accounts for surface heating due to, e.g., Ohmic decay of surrounding magnetic field, roto-chemical heating, vortex creep heating and crust cracking~\cite{Gonzalez:2010ta}.
While the emission of light particles contributes to the cooling with an extra negative term (see e.g.~\cite{Iwamoto:1984ir,Fiorillo:2025zzx,Fiorillo:2026dqu,Fiorillo:2026wso}), the proton conversion  processes~\eqref{pion0_process} and~\eqref{gamma_process} inject energy in the NS core in the form of photons and neutral pions, while~\eqref{np_process} and~\eqref{pp_process} inject energy in the form of positrons that subsequently annihilate with electrons $e^+e^-\to 2\gamma$. As the temperature decreases, such temperature-independent term will eventually dominate.

The total energy injection rate per unit volume can be computed by evaluating
the following quantity
\beq\label{eq:epsilon_NS}
\epsilon_{\rm tot} = \epsilon_{pe\to \gamma \phi} + \epsilon_{pe\to \pi\phi}+ \epsilon_{np\to e n\phi}+\epsilon_{pp\to ep\phi},
\eeq
where for each process $\rm P$ the energy injection rate calculation is carried out similarly as the rate $\gamma_{\rm P}$ as 
\beq 
\epsilon_{\rm P}=C_{m\to n}[|{\cal M}_{\rm P}|^2f_1\dots f_m (1\pm f_{m+1})\dots(1\pm f_{m+n}) E_{\rm inj}] . 
\eeq
Here $E_{\rm inj}$ is the energy of the final state particle that is not $\phi^*$ nor a nucleon. Thus, depending on the process, $E_{\rm inj}=\{E_\gamma, E_{\pi^0}, E_{e^+}+m_e\}$ (conservatively, the injected energy is the energy of the positron plus the mass of an electron that later annihilates with the positron).
We compute the energy injection rates for two NS density profiles and show their resulting radial dependence in the middle panel of Fig.~\ref{fig:rates_nB_r}. Integrating the energy injection rate over the NS density profile we obtain the heating rate 
\begin{align}
    H_{\rm tot}  = \int dr \ 4\pi r^2\epsilon_{\rm tot}(n_B(r))\ .
\end{align}

Thus, a stringent bound on the coupling $y$ is obtained by requiring the heating to not exceed the Stefan-Boltzmann luminosity. Taking the coldest NS measured to date, whose upper bound on the temperature is $T_{\rm NS}\lesssim 34100\,\rm K$~\cite{Guillot:2019ugf}, we require that $H_{\rm tot} \lesssim L_{\gamma}$: any additional heating source would result in a higher surface temperature~\cite{McKeen:2020oyr,McKeen:2021jbh}. We model the coldest NS with the $1.5\, M_\odot$ density profile shown in the bottom panel of Fig.~\ref{fig:rates_nB_r} obtained from the APR EOS with catalyzed crust via \texttt{NSCool} \cite{NSCool} and we obtain $y\lesssim5.2\times 10^{-23}$. Notice that this argument would need to be modified for couplings so large that the structure of the NS is affected; this happens for $1.5\,M_\odot$ at $y\gtrsim 6\times 10^{-19}$. However, these values of $y$ are already excluded by requiring heavier NS ($2M_\odot$) to be stable, as shown above.

\section{E.~Phase space integrals}\label{app:phase_space}
In this section we show a powerful method based on Monte Carlo techniques to compute generic integrals in the full phase space of the involved particles. These methods are useful to compute rates throughout this work and beyond. 
\subsection{In-vacuum phase space integrals}
We are interested in integrals of the type
\begin{align}
    C_{m\to n}[{\cal F}]= \int \left[ \prod_{i=1}^{n+m} d\Pi_i\right] (2\pi)^4 \delta^{(4)}\left(\sum_{i=1}^m P_i -\sum_{j=m+1}^{m+n} P_j \right) {\cal F}(P_i\cdot P_j,m_i,E_i,\mu_i,T),
\end{align}
where ${\cal F}$ is, for example, a generic function of scalar products, energies, chemical potentials and temperature or other scalar quantities. Depending on the number of particles in the process the number of invariant scalar products can vary. We follow a general Monte Carlo approach. Denote
\begin{align}
    P_X &= P_1+\dots +P_m - P_{m+1}- \dots -P_{m+n-2} \\
    P_Y &=P_{m+n-1}\\
    P_Z &= P_{m+n}
\end{align}
Note that $Y$ and $Z$ are actual particles, with their own masses. We immediately integrate out ${\bf p}_Z$ so that
\begin{align}
   C_{m\to n}[{\cal F}] =g_{m+n}\int \left[ \prod_{i=1}^{n+m-1} d\Pi_i\right] \frac{2\pi}{2 E_Z} \delta(E_X-E_Y-E_Z) {\cal F}\ .
\end{align}
From now on $E_Z=\sqrt{m_Z^2+|{\bf p}_X -{\bf p }_Y|^2}$. Next we integrate out $p_Y=|{\bf p}_Y|$ using the remaining Dirac delta. We define angles $\theta_Y$ and $\phi_Y$ in a polar frame with zenith aligned with ${\bf p}_X$ and call $c_Y\equiv \cos \theta_Y$. We obtain
\begin{align}
    C_{m\to n}[{\cal F}] =g_{m+n}g_{m+n-1} \int \left[ \prod_{i=1}^{n+m-2} d\Pi_i\right ]\frac{2\pi}{2 E_Z}\frac{p_Y^2 dc_Y d\phi_Y}{2E_Y (2\pi)^3} J^{-1} \Theta[Q^2-(m_Y+m_Z)^2] \Theta(\Delta) {\cal F},
\end{align}
where
\begin{align}
    J  &\equiv \left|\frac{\partial(E_Y +E_Z)}{\partial p_Y} \right|= \left| \frac{p_Y}{E_Y}+\frac{p_Y-p_Xc_Y}{E_Z}\right|,\\
    Q^2 &\equiv E_X^2-p_X^2,\\
    E_X &\equiv E_Y+E_Z,\\
    \Delta &\equiv m_Y^4 + (m_Z^2-Q^2)^2 - 2 m_Y^2
[Q^2+2(1-c_Y^2)p_X^2+m_Z^2] .
\end{align}
The condition $\Delta\geq0$ comes from the existence of a solution $p_Y$ for energy conservation $E_X=E_Y+E_Z$:
\begin{align}
    p_Y = \frac{c_Y p_X (Q^2-m_Z^2+m_Y^2) + E_X \sqrt{\Delta}}{2(E_X^2-c_Y^2p_X^2)}\ .
\end{align}
To make the Monte Carlo integration over a finite volume possible for thermally distributed particles, it is appropriate to perform a variable transformation of momenta $p_i =-\Lambda_i\log (x_i)$ with $dp_i= -\Lambda_i dx_i/x_i=j_i dx_i$. 
For degenerate particle species, it is more appropriate to instead generate uniformly $0\leq x_i\leq 1$ with $x_i=p_i/p_i^{\rm F}$ given the Fermi momentum of the species. The Jacobian of this transformation is then $dp_i = p_i^{\rm F}  dx_i=j_i dx_i$.
The strategy to evaluate the integral is to randomly generate triples $(x_i,c_i,\phi_i)$ for $i=1,\dots, m+n-2$ and construct their respective particles' four momenta $P_i=(E_i=\sqrt{m_i^2 +{\bf p}_i^2},{\bf p}_i)$ from their masses. Then one builds $P_X=\sum_{i=1}^{n+m-2}( E_i , {\bf p}_i)$ (with proper signs), generates $c_Y, \phi_Y$, derives $p_Y$ from the formula above and computes $P_Y=(E_Y,{\bf p}_Y)$. Finally one obtains $P_Z=P_X-P_Y$. Therefore in total $3(m+n-2)+2$ variables are generated. The integral is evaluated as the average over samples
\begin{align}
    C_{m\to n}[{\cal F}]\approx \frac{(4\pi)^{n+m-1}}{(2\pi)^{3(n+m-2)}}\bigg[\prod_{i=1}^{n+m}g_i\bigg]\bigg\langle \left[ \prod_{i=1}^{n+m-2} j_i \frac{p_i^2}{2E_i}\right] \frac{p_Y^2}{(2\pi)^3  2E_Y} \frac{2\pi}{2E_Z} J^{-1}\Theta(\Delta)\Theta[Q^2-(m_Y+m_Z)^2]{\cal F}\bigg\rangle \ .
\end{align}
\subsection{In-medium phase space integrals}
In a medium the phase space integrals are modified in several ways. Due to self-energy effects, for baryons, in-medium kinetic momenta $k_\mu ^*=k_\mu-\Sigma_\mu$ are employed in matrix elements and energies $E^* = E-\Sigma_0$ appear in the Lorentz invariant phase space measure instead of standard vacuum energies $E$. 

To allow \textit{any} particle $i$ to be treated this way, let's consider $\pmb \Sigma=0$ and \textit{all} the dispersion relations to be of the form $E_i=\sqrt{{m_i^*}^2+p_i^2}+\Sigma_0^i= E^*_i+\Sigma_0^i$ and ${\bf p}_i={\bf p}_i^*$. If the particle $i$ is not a baryon, simply we can take $\Sigma_0^i=0$ and $m^*_i=m_i$ so that $E_i=E_i^*$ follows. Note that $P_i^2 = {E_i}^2-p_i^2={m_i^*}^2+ {\Sigma_0^i}^2+2{E_i^*}\Sigma_0^i$.

We start from 
\begin{align}
    C_{m\to n}[{\cal F}]= \int \left[ \prod_{i=1}^{n+m} d\Pi_i^*\right] (2\pi)^4 \delta^{(4)}\left(\sum_{i=1}^m P_i -\sum_{j=m+1}^{m+n} P_j \right) {\cal F}(P_i^*\cdot P_j^*,m_i^*,E_i^*,\mu_i,T),
\end{align}
with $d\Pi_i^* = d^3p_i/(2E^*_i)/(2\pi)^3$.
We proceed similarly as in the vacuum case, integrating out ${\bf p}_Z$ using the Dirac delta
\begin{align}
   C_{m\to n}[{\cal F}] =g_{n+m}\int \left[ \prod_{i=1}^{n+m-1} d\Pi_i^*\right] \frac{2\pi}{2 E_Z^*} \delta(E_X-E_Y-E_Z) {\cal F}\ .
\end{align}
where now $E_Z=\sqrt{{m_Z^*}^2+|{\bf p}_X -{\bf p }_Y|^2}+\Sigma_0^Z$. Now we want to integrate out $p_Y$ using the remaining energy conservation delta. We obtain
\begin{align}
    C_{m\to n}[{\cal F}] = g_{n+m}g_{n+m-1}\int \left[ \prod_{i=1}^{n+m-2} d\Pi_i\right ]\frac{2\pi}{2 E_Z^*}\frac{p_Y^2 dc_Y d\phi_Y}{2E_Y^* (2\pi)^3} J^{-1} \Theta[Q^2-(m_Y^*+m_Z^*)^2] \Theta(\widetilde\Delta) {\cal F},
\end{align}
where
\begin{align}
    J&\equiv \left|\frac{\partial(E_Y +E_Z)}{\partial p_Y} \right|= \left| \frac{p_Y}{E_Y^*}+\frac{p_Y-p_Xc_Y}{E_Z^*}\right|,\\
   \widetilde Q^2 &\equiv \widetilde E_X^2-p_X^2,\\
    \widetilde E_X &\equiv E_Y+E_Z-\Sigma_0^Y-\Sigma_0^Z=\sqrt{{m_Y^*}^2+p_Y^2}+\sqrt{{m_Z^*}^2+p_X^2-2c_Y p_Xp_Y+p_Y^2},\\
    \widetilde \Delta &\equiv {m_Y^*}^4 + ({m_Z^*}^2-\widetilde Q^2)^2 - 2 {m_Y^*}^2
[\widetilde Q^2+2(1-c_Y^2)p_X^2+m_Z^2] .
\end{align}
The condition $\widetilde\Delta\geq0$ comes from the existence of a solution $p_Y$ for energy conservation $E_X=E_Y+E_Z$:
\begin{align}
    p_Y = \frac{c_Y p_X (\widetilde Q^2-{m_Z^*}^2+{m_Y^*}^2) + \widetilde E_X \sqrt{\widetilde\Delta}}{2(\widetilde E_X^2-c_Y^2p_X^2)}\ .
\end{align}
The difference with the in-vacuum case amounts to the substitution $E_X\to \widetilde E_X$ and $m_i\to m_i^*$.
As before one performs a transformation, depending on the most adequate sampling, for momenta $dp_i=j_i dx_i$. To evaluate the integral we randomly generate triples $(x_i,c_i,\phi_i)$ for $i=1,\dots, m+n-2$ and construct their respective particles' four momenta $P_i=(E_i=\sqrt{{m_i^*}^2 +{\bf p}_i^2}+\Sigma_0^i,{\bf p}_i)$ from their masses. Then one builds $P_X=\sum_{i=1}^{n+m-2}( E_i , {\bf p}_i)$ (with proper signs), generates $c_Y, \phi_Y$, derives $p_Y$ from the formula above and computes $P_Y=(E_Y,{\bf p}_Y)$. Finally one obtains $P_Z=P_X-P_Y$. Therefore in total $3(m+n-2)+2$ variables are generated. The integral is evaluated as the average over samples:
\begin{align}
    C_{m\to n}[{\cal F}]\approx \frac{(4\pi)^{n+m-1}}{(2\pi)^{3(n+m-2)}}\bigg[\prod_{i=1}^{n+m}g_i\bigg]\bigg\langle \left[ \prod_{i=1}^{n+m-2} j_i \frac{p_i^2}{2E_i^*}\right] \frac{p_Y^2}{(2\pi)^3  2E_Y^*} \frac{2\pi}{2E_Z^*} J^{-1}\Theta(\widetilde \Delta)\Theta[\widetilde Q^2-(m_Y^*+m_Z^*)^2]{\cal F}\bigg\rangle \ .
\end{align}
%TC:endignore

\end{document}